\documentclass[twocolumn]{aastex63}

\hypersetup{pdfnewwindow=true,
     colorlinks=true,
     linkcolor=blue,
     citecolor=blue,
	 final=true}
	 
\usepackage{etoolbox}
\makeatletter
\patchcmd{\NAT@citex}
  {\@citea\NAT@hyper@{%
     \NAT@nmfmt{\NAT@nm}%
     \hyper@natlinkbreak{\NAT@aysep\NAT@spacechar}{\@citeb\@extra@b@citeb}%
     \NAT@date}}
  {\@citea\NAT@nmfmt{\NAT@nm}%
   \NAT@aysep\NAT@spacechar\NAT@hyper@{\NAT@date}}{}{}
\patchcmd{\NAT@citex}
  {\@citea\NAT@hyper@{%
     \NAT@nmfmt{\NAT@nm}%
     \hyper@natlinkbreak{\NAT@spacechar\NAT@@open\if*#1*\else#1\NAT@spacechar\fi}%
       {\@citeb\@extra@b@citeb}%
     \NAT@date}}
  {\@citea\NAT@nmfmt{\NAT@nm}%
   \NAT@spacechar\NAT@@open\if*#1*\else#1\NAT@spacechar\fi\NAT@hyper@{\NAT@date}}
  {}{}
\makeatother

\usepackage{color,comment}
\usepackage[normalem]{ulem}

\newcommand{\simba}{\textsc{simba}}
\newcommand{\pd}{\textsc{powderday}}
\newcommand{\Msun}{\mathrm{M}_{\sun}} 

\received{26 May 2021}
\revised{7 March 2022}
\accepted{11 March 2022}
\submitjournal{ApJ}

\shorttitle{The UVJ Diagram in SIMBA}
\shortauthors{Akins et al.}

\begin{document}

\title{\large Quenching and the UVJ diagram in the SIMBA cosmological simulation}

\correspondingauthor{Hollis B. Akins}
\email{hollis.akins@gmail.com}

\author[0000-0003-3596-8794]{Hollis B. Akins}
\affiliation{Department of Physics, Grinnell College, 1116 Eighth Ave., Grinnell, IA 50112, USA}
\author[0000-0002-7064-4309]{Desika Narayanan}
\affiliation{Department of Astronomy, University of Florida, 211 Bryant Space Sciences Center, Gainesville, FL 32611, USA}
\affiliation{University of Florida Informatics Institute, 432 Newell Drive, CISE Bldg. E251, Gainesville, FL 32611, USA}
\affiliation{Cosmic Dawn Center (DAWN), Niels Bohr Institute, University of Copenhagen, Juliane Maries vej 30, DK-2100 Copenhagen, Denmark}
\author[0000-0001-7160-3632]{Katherine E. Whitaker}
\affiliation{Department of Astronomy, University of Massachusetts, Amherst, MA 01003, USA}
\affiliation{Cosmic Dawn Center (DAWN), Niels Bohr Institute, University of Copenhagen, Juliane Maries vej 30, DK-2100 Copenhagen, Denmark}
\author[0000-0003-2842-9434]{Romeel Davé}
\affiliation{Institute for Astronomy, Royal Observatory, University of Edinburgh, Edinburgh, EH9 3HJ, UK}
\affiliation{University of the Western Cape, Bellville, Cape Town 7535, South Africa}
\affiliation{South African Astronomical Observatory, Observatory, Cape Town 7925, South Africa}
\author[0000-0003-4422-8595]{Sidney Lower}
\affiliation{Department of Astronomy, University of Florida, 211 Bryant Space Sciences Center, Gainesville, FL 32611, USA}
\author[0000-0001-5063-8254]{Rachel Bezanson}
\affiliation{Department of Physics and Astronomy and PITT PACC, University of Pittsburgh, Pittsburgh, PA 15260, USA}
\author[0000-0002-1109-1919]{Robert Feldmann}
\affiliation{Institute for Computational Science, University of Zurich, CH-8057 Zurich, Switzerland}
\author[0000-0002-7613-9872]{Mariska Kriek}
\affiliation{Department of Astronomy, University of California, Berkeley, CA 94720, USA}

\begin{abstract}
Over the past decade, rest-frame color-color diagrams have become popular tools for selecting quiescent galaxies at high redshift, breaking the color degeneracy between quiescent and dust-reddened star-forming galaxies. 
In this work, we study one such color-color selection tool---the rest-frame $U-V$ vs. $V-J$ diagram---by employing mock observations of cosmological galaxy formation simulations. 
In particular, we conduct numerical experiments assessing both trends in galaxy properties in UVJ space and the color-color evolution of massive galaxies as they quench at redshifts $z\sim 1$--$2$. 
We find that our models broadly reproduce the observed UVJ diagram at $z=1$--$2$, including (for the first time in a cosmological simulation) reproducing the population of extremely dust-reddened galaxies in the top right of the UVJ diagram.
However, our models primarily populate this region with low-mass galaxies and do not produce as clear a bimodality between star-forming and quiescent galaxies as is seen in observations. 
The former issue is due to an excess of dust in low-mass galaxies and relatively gray attenuation curves in high-mass galaxies, while the latter is due to the overpopulation of the green valley in \simba. 
When investigating the time evolution of galaxies on the UVJ diagram, we find that the quenching pathway on the UVJ diagram is independent of the quenching timescale, and instead dependent primarily on the average specific star formation rate in the 1 Gyr prior to the onset of quenching. 
Our results support the interpretation of different quenching pathways as corresponding to the divergent evolution of post-starburst and green valley galaxies.
\end{abstract}

\keywords{\small \href{http://astrothesaurus.org/uat/1724}{Two-color diagrams (1724)}; \href{http://astrothesaurus.org/uat/2040}{Galaxy quenching (2040)}; \href{http://astrothesaurus.org/uat/2176}{Post-starburst galaxies (2176)}}

\section{Introduction} \label{sec:introduction}

Understanding and quantifying the rate of star-formation at high redshift is key to constraining the formation of massive galaxies in the early universe. 
It has been widely observed that massive galaxies generally fall into two categories: blue, disk-dominated galaxies on the star-forming main sequence (SFMS), and red, elliptical, quiescent galaxies \citep{strateva_color_2001, baldry_quantifying_2004, balogh_bimodal_2004, bell_nearly_2004, faber_galaxy_2007}.
While quiescent galaxies are ubiquitous in the local universe, recent observations have detected massive quiescent galaxies out to $z \sim 4$ \citep{glazebrook_massive_2017,schreiber_infrared_2018a, carnall_timing_2020a,forrest_extremely_2020, valentino_quiescent_2020}.
However, it can be difficult to identify quiescent galaxies and constrain their SFRs at high redshift owing to the ubiquity of dust-obscured star-formation at $z\gtrsim 1$, which can significantly redden star-forming galaxies \citep[SFGs][]{brammer_dead_2009, maller_intrinsic_2009}. 

Over the past decade, rest-frame color-color diagrams have become popular tools for breaking this degeneracy between SFGs reddened by dust and quiescent galaxies, intrinsically red due to older stellar populations. 
Such diagrams typically compare one color in the rest-frame near-UV (NUV) to optical range and another in the rest-frame optical--near-IR in order to cleanly separate quiescent and dusty SFGs galaxies on the optical red sequence. 
While spectroscopic measures such as the H$\alpha$ luminosity and D$_n(4000)$ index can serve as more reliable indicators of active star-formation \citep[e.g.][]{kauffmann_stellar_2003}, color-color diagrams can be readily applied to large surveys and at high redshift \citep[e.g.][]{daddi_new_2004, labbe_irac_2005, wuyts_what_2007, arnouts_swirevvdscfhtls_2007, williams_detection_2009, ilbert_mass_2013, tomczak_galaxy_2014, kriek_mosfire_2015, wu_fast_2018, fang_demographics_2018, carnall_vandels_2019}. 
In particular, the rest-frame $U-V$ vs.~$V-J$ (hereafter UVJ) diagram has proven an effective diagnostic for selecting quiescent galaxies across a range of redshifts \citep{wuyts_what_2007, williams_detection_2009, whitaker_newfirm_2011, muzzin_evolution_2013, fang_demographics_2018}.
In addition to providing an accessible method for selecting quiescent galaxies, UVJ colors have been shown to correlate with specific star-formation rates \citep[sSFRs;][]{williams_evolving_2010, patel_starformationratedensity_2011,leja_uvj_2019}, dust attenuation \citep[$A_V$;][]{price_direct_2014, forrest_uv_2016, martis_evolution_2016, fang_demographics_2018}, and stellar age \citep{whitaker_quiescent_2013, belli_mosfire_2019, carnall_vandels_2019}.

Despite its central role as a selection tool for high-redshift quiescent galaxies, much is still uncertain about the distribution of galaxy properties on the UVJ diagram. 
In particular, the inferred properties and interpretations of galaxy positions in UVJ space may be sensitive to the assumed dust attenuation curve. 
Often, at high redshift, all galaxies are assumed to follow a \citet{calzetti_dust_2000} dust attenuation law, though recent evidence points to the likelihood that galaxies span a range of attenuation curves \citep{kriek_dust_2013, scoville_dust_2015, salmon_breaking_2016, leja_deriving_2017,salim_dust_2018, narayanan_theory_2018, salim_dust_2020a}. 
While slopes and feature strengths of attenuation curves naturally correlate with galaxy properties such as SFR and $M_*$ \citep{salim_dust_2018}, much of the variation in attenuation curves seems to be driven by less constrained factors such as the complexity of the relative star-dust geometry \citep{seon_radiative_2016, narayanan_theory_2018, trayford_fade_2020}. 
Indeed, the spread of galaxy colors in the star-forming region of UVJ space has been shown to be correlated with galaxy morphology and inclination \citep{patel_uvj_2012, zuckerman_reproducing_2021}, and variations in the attenuation curve have been hypothesized to lead to UVJ misidentification \citep{roebuck_simulations_2019}.
While the UVJ diagram has proven an effective tool, there is still a great deal of uncertainty regarding the utility of color-color diagrams in inferring galaxy properties, and independent measures of such properties are necessary to resolve this. 

Furthermore, the ubiquity of the UVJ diagram as a selection and visualization tool at high-redshift has sparked interest in how different galaxy evolutionary histories (i.e., different quenching mechanisms or timescales) manifest in UVJ space.
For example, recently quenched post-starburst galaxies have been observed to cluster in a unique region of UVJ space \citep{wild_evolution_2016, whitaker_large_2012, yano_relation_2016, almaini_massive_2017, suess_color_2020}.
Similarly, \citet{fang_demographics_2018} identify a population of ``transition'' galaxies in the star-forming region of UVJ space but with suppressed SFRs, and propose that the mass distribution of these transition galaxies implies a mass-dependent quenching path in UVJ space. 
Some authors have inferred the UVJ evolutionary tracks for galaxies based on their SFHs and modeling a relationship between SFR and dust attenuation \citep[e.g.][]{barro_candels_2014, belli_mosfire_2019, carnall_vandels_2019, suess_dissecting_2021}.
These model tracks support the view of an evolutionary pathway dependent upon the quenching mechanism, in which faster-quenching post-starburst galaxies enter the quenched region from the bottom left and slower-quenching galaxies enter from the right \citep[see e.g.][Figure 12]{suess_dissecting_2021}. 
However, these models are highly dependent upon the assumed relationship between dust attenuation and SFR, which is unconstrained for galaxies at the epoch of quenching and may not be universal. 
A more complete and consistent theory for the evolution of galaxies in color-color space, though elusive, may provide efficient selection methods for studies of particular quenching processes. 

In this light, cosmological simulations can help us understand and contextualize the distribution and evolution of galaxies on the UVJ diagram, as they provide easy access to fundamental galaxy properties over time. 
The UVJ selection technique has been explored in theoretical work in the past, and observations of the UVJ diagram have been broadly reproduced in cosmological \citep[e.g.][]{dave_mufasa_2017, donnari_star_2019}, zoom-in \citep[e.g.][]{feldmann_colours_2017}, and idealized \citep[e.g.][]{roebuck_simulations_2019} galaxy evolution simulations. 
As of yet, however, there has been no fully cosmological model that employs both realistic models of dust (to attend to the aforementioned issues of dust obscuration and attenuation) and radiative transfer (to model the mock colors) to thereby explore galaxies in UVJ space.  
The purpose of this paper is to develop and explore such a model.

In this work, we examine trends on the UVJ diagram using the \simba\ suite of simulations \citep{dave_simba_2019}, and using the 3D dust radiative transfer code \pd\ \citep{narayanan_powderday_2021}. 
The structure of the paper is as follows.
In Section~\ref{sec:methods} we describe the \simba\ simulations, the \pd\ dust radiative transfer code, and outline our fiducial definitions. 
In Section~\ref{sec:uvj_observations} we compare our model UVJ diagram to observations both with respect to the distribution of UVJ colors (\ref{sec:uvj_obs_dist}) and trends in galaxy properties in UVJ space (\ref{sec:uvj_trends}).
This has the primary purpose of interrogating the simulations' ability to reproduce observations.
In Section~\ref{sec:evolution} we study the time evolution of galaxies in UVJ space with particular attention to the different pathways for quenching.
We compare our models to those employed in other theoretical work in Section~\ref{sec:discussion}, and we summarize our conclusions in Section~\ref{sec:conclusions}. 

Throughout this paper, we adopt a \citet{kroupa_initial_2002} initial mass function (IMF) and a cosmology consistent with the Planck Collaboration (\citeyear{planckcollaboration_planck_2016a}): $\Omega_m = 0.3$, $\Omega_{\Lambda} = 0.7$, $\Omega_b = 0.048$, $H_0 = 68~\mathrm{km}~\mathrm{s}^{-1}~\mathrm{Mpc}^{-1}~h^{-1}$, $\sigma_8 = 0.82$, and $n_s = 0.97$.

\section{Methods} \label{sec:methods}

\subsection{Simulations}\label{sec:simulations}

This work utilizes the \simba\ simulations, a series of state-of-the-art cosmological hydrodynamic simulations of galaxy formation \citep{dave_simba_2019}. 
The \simba\ simulations are the successor to the \textsc{mufasa} \citep{dave_mufasa_2016} simulations, and are run using a modified version of the gravity plus hydrodynamics solver \textsc{gizmo} \citep{hopkins_new_2015}, which uses the \textsc{gadget-3} tree-particle-mesh gravity solver \citep{springel_cosmological_2005} and a meshless finite-mass method for hydrodynamics. 
A detailed description of the simulation physics and methodology has been presented in \citet{dave_simba_2019}. 
We refer the reader to this work for details and summarize the salient points here. 

\simba\ models star-formation using a molecular hydrogen (H$_2$)-based \citet{schmidt_rate_1959} relation, where the H$_2$ fraction is computed using the subresolution model of \citet{krumholz_comparison_2011} based on the metallicity and local column density, with minor modifications as described in \citet{dave_mufasa_2016} to account for numerical resolution. 
The instantaneous star-formation rate is thus given by the H$_2$ density divided by the dynamical time: $\mathrm{SFR} = \epsilon_* \rho_{\mathrm{H2}} / t_{\mathrm{dyn}}$, where we use $\epsilon_* = 0.02$ \citep{kennicutt_global_1998}.
Radiative cooling and photoionization heating are modeled using the \textsc{grackle-3.1} library \citep{smith_grackle_2017}, including metal cooling and non-equilibrium evolution of primordial elements. 
The chemical enrichment model tracks eleven metals during the simulation, with enrichment tracked from Type II supernovae (SNe), Type Ia SNe, and asymptotic giant branch (AGB) stars. 
Star formation-driven galactic winds are modeled as decoupled two-phase winds, with 30\% of wind particles ejected ``hot,'' and with a mass loading factor that scales with stellar mass, based on the Feedback In Realistic Environments (FIRE) \citep{hopkins_galaxies_2014} zoom simulation scalings from \citet{angles-alcazar_cosmic_2017}. 

\simba\ builds upon \textsc{mufasa} through the addition of black hole growth via torque-limited accretion 
\citep{hopkins_analytic_2011, angles-alcazar_black_2013, angles-alcazar_torquelimited_2015} and AGN feedback via bipolar kinetic outflows. 
Black holes are seeded and grown during the simulation, and the accretion energy drives feedback that acts to quench galaxies. 
For cold gas ($T<10^5$ K), black hole growth is implemented following the torque-limited accretion model of \citet{angles-alcazar_gravitational_2017} which is based on \citet{hopkins_analytic_2011}, while for hot gas ($T > 10^5$ K) Bondi accretion \citep{bondi_spherically_1952} is adopted. 
AGN feedback is implemented with a model designed to mimic the observed dichotomy of black hole growth and feedback modes observed \citep[e.g.][]{heckman_coevolution_2014}. 
In particular, real AGN show a ``radiative'' mode at high Eddington ratios ($f_{\mathrm{Edd}}$) characterized by mass-loaded radiatively driven winds and a ``jet'' mode at low $f_{\mathrm{Edd}}$, characterized by high velocity jets of $\sim 10^4$ km s$^{-1}$. 
The AGN outflow model has three modes of feedback: radiative, jet, and X-ray. 
Radiative and jet modes are implemented kinetically, with outflows ejected following a variable velocity and mass outflow rate to mimic the transition between high mass-loaded radiative winds and high-velocity jets. 
Full velocity jets are achieved at low Eddington ratios ($f_{\mathrm{Edd}} < 0.02$) and high black hole masses ($M_{\mathrm{BH}} > 10^{7.5}~\Msun$).
X-ray feedback directly increases the temperature of non-ISM gas and both heats and expels ISM gas. 
As shown in \citet{dave_simba_2019}, the jet mode is primarily responsible for quenching galaxies, while X-ray feedback has an important role in suppressing residual star formation. 

Of particular relevance for this work, \simba\ includes a unique self-consistent on-the-fly subgrid model for the production, growth, and evolution of dust grains \citep[described in detail in][]{dave_simba_2019, li_dusttogas_2019}. 
Dust grains are assumed to have a single size of $0.1~\mu$m and are passively advected with gas elements as a fraction of the element's metal budget. 
Dust grains grow via condensation following \citet{dwek_evolution_1998} but with updated condensation efficiencies, as well as by accretion of gas-phase metals via two-body collisions. 
Dust is destroyed (returned back to the gaseous metal phase) by collisions with thermally excited gas following the analytic approximation of dust growth rates from \citet{tsai_interstellar_1995}. 
A mechanism for dust destruction via SN shocks is implemented following \citet{mckinnon_dust_2016}. 
Dust is instantaneously destroyed in hot winds, during star-formation, and in gas impacted by jet or X-ray AGN feedback; however, dust is not destroyed in cold SF winds or radiative-mode AGN feedback to allow these winds to transport dust out of the galaxy. 
This model results in dust-to-metal ratios and dust mass functions in good agreement with observations for star-forming galaxies \citep{li_dusttogas_2019}. 

\begin{figure*}
    \centering
    \includegraphics[width=\linewidth]{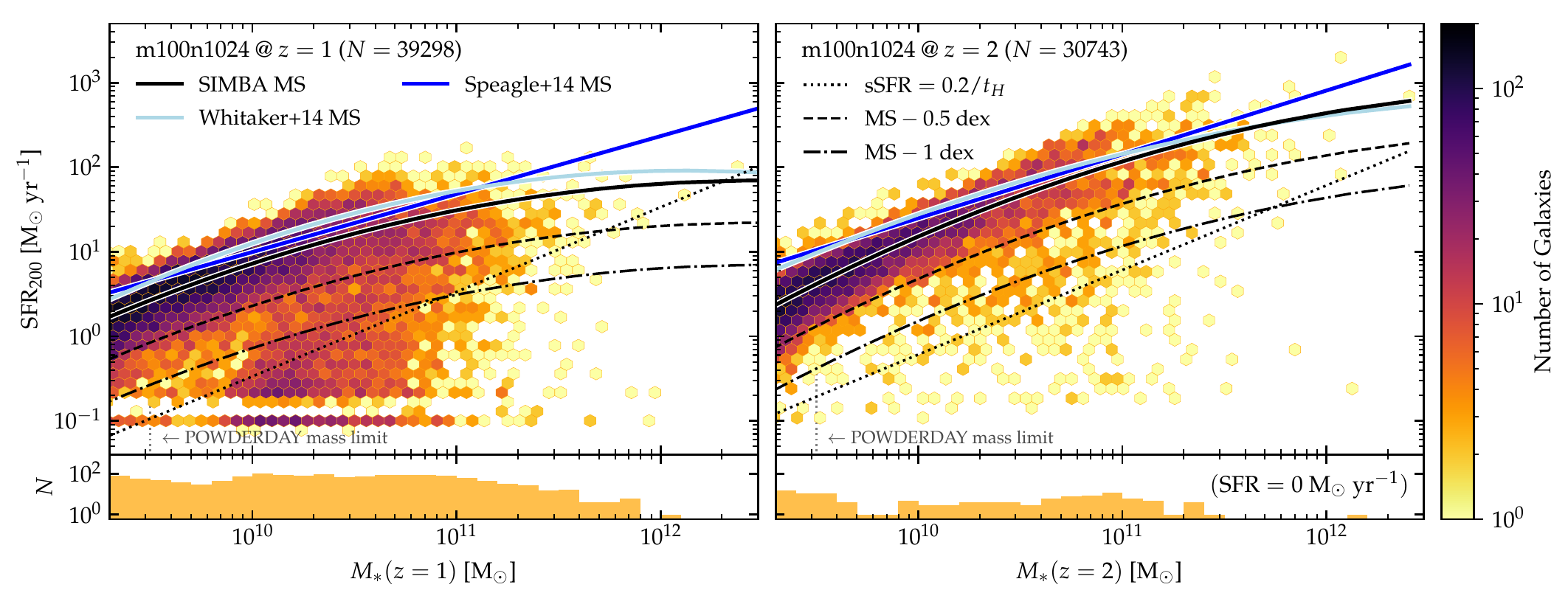}
    \caption{
    Hexbin plots of the star formation rate--stellar mass relation in our simulations at $z=1$ (left panel) and $z=2$ (right panel). The SFRs are averaged over the past $200$ Myr, and the solid black line shows the best-fit 2nd-degree polynomial for the \simba\ main sequence. 
    The full sample of $>30,000$ resolved \simba\ galaxies is used to determine the main sequence.
    The dark and light blue lines show the MS relations from \citet{speagle_highly_2014} and \citet{whitaker_constraining_2014}, respectively. 
    The dashed and dot-dash lines show the \simba\ 	$\mathrm{MS}-0.5$ dex and $\mathrm{MS}-1$ dex lines, respectively, which we use to separate between star-forming, transitioning, and quenched galaxies. 
    A simple sSFR cut of $\mathrm{sSFR} = 0.2/t_H$, where $t_H$ is the Hubble time at that redshift, is also shown as a dotted line. 
    The histogram on the lower panel shows the distribution of galaxies with no resolved star formation in the past 200 Myr, and the distinct horizontal feature at $\text{SFR} \sim 10^{-1}~M_\odot~\text{yr}^{-1}$ corresponds to the smallest resolvable SFR.}
    \label{fig:MS}
\end{figure*}

The primary simulation we use in this work is the fiducial $(100~\mathrm{Mpc}~h^{-1})^3$ comoving volume, run from $z=249$ to $z=0$ with $1024^3$ gas elements and $1024^3$ dark matter particles. 
The minimum gravitational softening length is $\epsilon_{\mathrm{min}} = 0.5~\mathrm{kpc}~h^{-1}$ and the mass resolution is $9.6\times 10^7~\Msun$ for dark matter particles and $1.8\times10^7~\Msun$ for gas elements. 
This simulation outputs 151 snapshots from $z=20 \to 0$. 
We supplement our analysis with results from the high-resolution $(25~\mathrm{Mpc}~h^{-1})^3$ comoving box. 
This run includes $512^3$ gas elements and $512^3$ dark matter particles, with a mass resolution of $1.2\times10^7$ and $2.3\times 10^6~\Msun$ for dark matter and gas elements, respectively. 
In addition to eight-times higher mass resolution, the run outputs twice as many snapshots, for a total of 305 from $z=20 \to 0$. 
This increased time resolution is the primary reason we include this simulation in this work; however, as a bonus, this provides a view of low-mass galaxies and serves as a test of numerical convergence. 
Unless otherwise stated, results are drawn from the $100~\mathrm{Mpc}~h^{-1}$ box. 

Galaxy properties are computed and cataloged using \textsc{caesar},\footnote{Available at \url{https://github.com/dnarayanan/caesar}} an extension of the \textsc{yt} simulation analysis software \citep{turk_yt_2011}. 
\textsc{caesar} identifies galaxies using a 6D friends-of-friends algorithm with a spatial linking length of 0.0056 times the mean interparticle separation and a velocity linking length set to the local velocity dispersion.
\textsc{caesar} outputs a cross-matched halo and galaxy catalog, from which the bulk of galaxy properties used in this work are drawn. 
Additionally, \textsc{caesar} includes a progenitor/descendant tracking module that identifies the major progenitor and descendant of a given galaxy at a given snapshot based on the number of star particles in common. 
We utilize this code to track galaxies across snapshots.

\subsection{Fiducial Definitions} \label{sec:definitions}

A consistent challenge for studies of the shutoff of SF in galaxies is that there exists no standardized, widely accepted definition of ``quenching.'' 
Here, we present the \simba\ SFR-$M_*$ relation, from which we establish our fiducial definition of quenching and compare to other common definitions. 

Figure~\ref{fig:MS} shows the $\mathrm{SFR}$-$M_*$ relation for \simba\ at $z=1$ and $z=2$.
Throughout this work, we compute star-formation rates by summing (and normalizing) the formation masses of star particles formed over the past 200 Myr.\footnote{We use an averaging timescale of 200 Myr to balance the utility of instantaneous SFRs with the resolution of the simulations. For a thorough discussion, see Appendix A of \citet{donnari_star_2019}.}
Following \citet{whitaker_constraining_2014}, we adopt a ``bending'' model for the star-forming main sequence and fit a 2nd-order polynomial to the running median of $\log {\rm SFR}$ in 0.2 dex bins of $\log M_*/\Msun$. 
We perform this fit iteratively, each time limiting the next fit to only star-forming galaxies with SFRs within 0.5 dex of the main sequence line computed in the previous iteration.
We compute best-fit coefficients \citep[labeled following Equation 2 of ][]{whitaker_constraining_2014} of 
$a = -25.02$, $b = 4.19$, and $c = -0.16$ at $z=2$, and $a = -22.84$, $b = 3.95$, and $c = -0.16$ at $z=1$.
Figure~\ref{fig:MS} shows our MS fit alongside observational estimates for the SFMS from \citet{speagle_highly_2014} and \citet{whitaker_constraining_2014}.
We find that our SFMS fit is in good agreement with observations, though notably is lower in amplitude by $\sim 0.3$ dex \citep[as in][]{dave_simba_2019, nelson_spatially_2021}. 

We categorize galaxies based on their distance, in dex, from the \simba\ main sequence ($\Delta$SFR). 
If $\Delta\mathrm{SFR} < -1$ dex, we define the galaxy as quenched; if $\Delta\mathrm{SFR} > -0.5$ dex, we consider if on (or above) the SFMS. 
These two dividing lines are shown as dash-dot and dashed lines on Figure~\ref{fig:MS}, respectively.
If $\Delta\mathrm{SFR}$ is between these values, we consider the galaxy to be ``transitioning'' between the two populations. 
These fiducial definitions give a quenched fraction (for $M_* > 10^{10}~\Msun$) of $12\%$ at $z=2$ and $40\%$ by $z=1$. 
Of the ``transitioning'' population, $\sim 35\%$ are rejuvenating (i.e., they have ${\rm SFR}_{50\,{\rm Myr}} > {\rm SFR}_{200\,{\rm Myr}}$) at both $z=1$ and $z=2$.

While we adopt this as our fiducial definition, the lack of a standardized definition of quenching makes it necessary to compare this choice to others. 
Figure~\ref{fig:MS} also shows a time-evolving cut in the specific SFR ($\mathrm{sSFR} = \mathrm{SFR}/M_*$) commonly used to define quenching \citep[e.g.][]{pacifici_evolution_2016, rodriguezmontero_mergers_2019}. 
Additionally, though not shown, we explore a cut in the normalized SFR (nSFR), the ratio of a galaxy's current SFR to its lifetime average SFR. 
We find the $\mathrm{nSFR} = 0.1$ cut adopted by \citet{carnall_inferring_2018} is in good agreement with the specific SFR cut shown in Figure~\ref{fig:MS}, and both of these definitions broadly agree with our $\mathrm{MS}-1~\mathrm{dex}$ cut.
Though our $\Delta$SFR definition of quenching is more lenient (i.e., includes more quenched galaxies) at low masses, and more strict at higher masses, our results are not sensitive to this definition. 

\subsection{3D Dust Radiative Transfer}\label{sec:powderday}

To extract observables from the simulation we use the 3D dust radiative transfer code \pd.\footnote{Available at \url{https://github.com/dnarayanan/powderday}}
\pd\ provides a convenient, modular, and parallelizable framework for computing the dust-attenuated spectral energy distributions (SEDs) of galaxies in cosmological simulations. 
Fundamentally, the code weaves together \textsc{fsps} \citep{conroy_propagation_2010, conroy_propagation_2010a} for stellar population synthesis, \textsc{hyperion} \citep{robitaille_hyperion_2011} for Monte Carlo radiative transfer, and \textsc{yt} \citep{turk_yt_2011} for interfacing with cosmological simulation data. 
\pd\ is described in detail in \citet{narayanan_powderday_2021}; here  we summarize the relevant points.

\begin{figure*}
	\centering
	\includegraphics[width=0.87\linewidth]{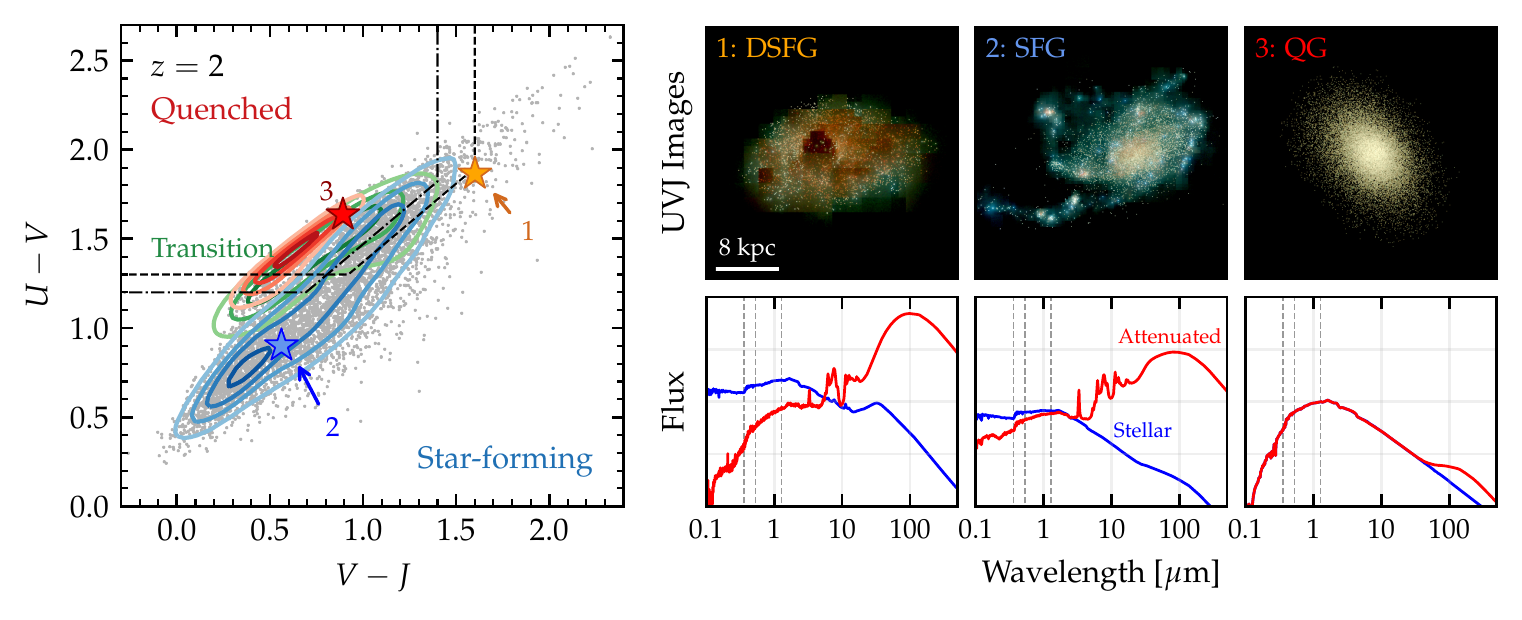}	
	\caption{Overview of the UVJ diagram derived from \pd\ radiative transfer. 
	\textit{Left:} The \simba\ UVJ diagram at $z=2$. KDE contours for the quenched galaxies, transitioning galaxies, and SFGs are plotted in shades of red, green, and blue, respectively, and our galaxy sample is plotted in the background. 
	The dashed and dashed-dotted lines show the $z=2$ UVJ selection criteria of \citet{williams_detection_2009} and \citet{whitaker_newfirm_2011}, respectively. 
	Stars indicate the UVJ colors of three example galaxies: 1) a DSFG, 2) a nondusty SFG, and 3) a quenched galaxy (QG). 
	\textit{Top right:} Simulated UVJ images of these three example galaxies derived from \pd. Particularly star-forming and particularly dusty regions can be seen in blue ($U$-band) and red ($J$-band), respectively. 
	\textit{Bottom right:} Stellar and dust-attenuated SEDs for these three example galaxies. 
	The central wavelengths of the U, V, and J bands are indicated as dashed lines behind the SEDs. }
	\label{fig:uvj_schematic}
\end{figure*}

For each galaxy identified by \textsc{caesar}, we perform stellar population synthesis using \textsc{fsps} \citep{conroy_propagation_2010, conroy_propagation_2010a}. 
We treat each star particle as a simple stellar population (SSP) with a fixed age and metallicity taken directly from the simulation. 
These properties are then provided to \textsc{fsps}, which generates a stellar SED assuming an initial mass function (IMF) combined with theoretical isochrones. 
We adopt MIST isochrones \citep{choi_mesa_2016, dotter_mesa_2016, paxton_modules_2011} and a MILES stellar spectra library \citep{sanchez-blazquez_mediumresolution_2006} as our fiducial choice of SPS parameters. 
We explore in Section~\ref{sec:uvj_obs_sps} the impact of the assumed isochrones on the resulting UVJ diagram.
Stellar SEDs for three example galaxies are shown as blue lines in the bottom right panels in Figure~\ref{fig:uvj_schematic}. 

We then compute the attenuated SEDs by performing dust radiative transfer. 
The dust properties stored in the gas elements in the simulation are projected on an adaptive octree grid. 
We then allow radiation from sources to propagate through the dusty ISM of the galaxy, which acts to scatter, absorb, and re-emit incident radiation. 
This is done in a Monte Carlo fashion in \textsc{hyperion} \citep{robitaille_hyperion_2011}.
Photon packets are released with random direction and frequency and propagate until they escape the grid or reach some limiting optical depth, and an iterative procedure is used to calculate the equilibrium dust temperature. 
The output SEDs are then calculated through ray-tracing; such SEDs are shown as red lines in Figure~\ref{fig:uvj_schematic}. 
In this work, the viewing angle for ray-tracing is fixed relative to the coordinate system of the cosmological box. 
Though the viewing angle is a flexible parameter, fixing it this way means the observed inclination of a given galaxy is effectively random, as in observations.

While we adopt \pd\ as our fiducial method for modeling dust attenuation, we compare the resulting UVJ diagrams to those derived from other modeling approaches in Section~\ref{sec:uvj_obs_attenuation}. 
We note that the combination of the \simba\ explicit dust model and \pd\ radiative transfer has had great success in reproducing observations of dusty galaxies, including matching the observed number density of high-redshift submillimeter galaxies \citep{lovell_reproducing_2021}, as well as the observed dust-to-gas and dust-to-metals ratios at low and high redshift \citep{li_dusttogas_2019}.

\subsection{Sample Selection and Photometry}

Though we use the full sample of resolved \simba\ galaxies to define the MS, it would be computationally intractable to run radiative transfer on this full sample. 
Instead, we select galaxies with $\log M_*/\Msun > 9.5$ from the $\left(100~h^{-1} \mathrm{Mpc} \right)^3$ simulation, which corresponds to $\sim 250$ star particles and $\sim 500$ gas elements.  
In this work, we focus primarily on the \simba\ snapshots at $z=2$ and $z=1$, as these span the range of redshifts at which UVJ selection is most often used. 
At $z=2$ our mass-limited sample contains 5,795 galaxies, while at $z=1$ it contains 12,035 galaxies.

We compute the rest-frame $U$, $V$, and $J$ magnitudes by convolving the SED generated by \pd\ with the corresponding filter transmission curve. 
Specifically, we use the \citet{bessell_ubvri_1990} $U$ and $V$ curves and the Mauna Kea UKIRT WFCAM curve for $J$ \citep{hewett_ukirt_2006}.
While we do not apply any apertures to these photometric measurements, we have verified that the resulting UVJ diagram is largely unchanged by the use of $0.7''$ radius apertures as in the 3D-HST survey \citep{skelton_3dhst_2014}.
Additionally, we show in Appendix~\ref{appendix:noise} the model UVJ diagram with the addition of mock observational noise; specifically, the noise resulting from $\sim 5\%$ uncertainty in the photometric redshift. 
This noise scatters galaxies in all directions, producing a UVJ diagram that, qualitatively, more closely resembles observations. 
However, as this noise may obscure the trends of interest in our work, we do not include this in our fiducial model.

\section{Observational Comparisons}\label{sec:uvj_observations}

Before exploring the evolution of simulated galaxies in UVJ space, we perform comparisons to observations to determine whether our simulations and radiative transfer methodology can adequately reproduce observations, and where they can not, diagnose the underlying issues. 

First, we present in Figure~\ref{fig:uvj_schematic} our fiducial UVJ diagram at $z=2$.
Here, we denote the distribution of star-forming, transitioning, and quenched galaxies via contours. 
The right panels show SEDs and simulated UVJ images for three example galaxies drawn from the simulation box: a dusty star-forming galaxy, a non-dusty star-forming galaxy, and a quenched galaxy.
We broadly reproduce the observed distribution of UVJ colors, in that we see distinct clustering of quiescent galaxies along the top left, a star-forming sequence along the bottom-right, and transition galaxies in between. 

However, our models are in tension with observations in two notable ways. 
First, we do not produce as clear of a bimodal distribution of galaxies as is observed; that is, we see more overlap between the quiescent and star-forming populations.  
Second, our colors are generally bluer than observations, with a significant number of DSFGs populating the quiescent region as defined by \citet{williams_detection_2009} and \citet{whitaker_newfirm_2011}. 
We explore these discrepancies further in the following subsections, first by focusing only on the distribution of UVJ colors and then by examining trends in galaxy properties on the UVJ diagram. 

\subsection{Distribution of UVJ Colors}\label{sec:uvj_obs_dist} 

\begin{figure*}
	\centering
	\includegraphics[width=
	0.87\linewidth]{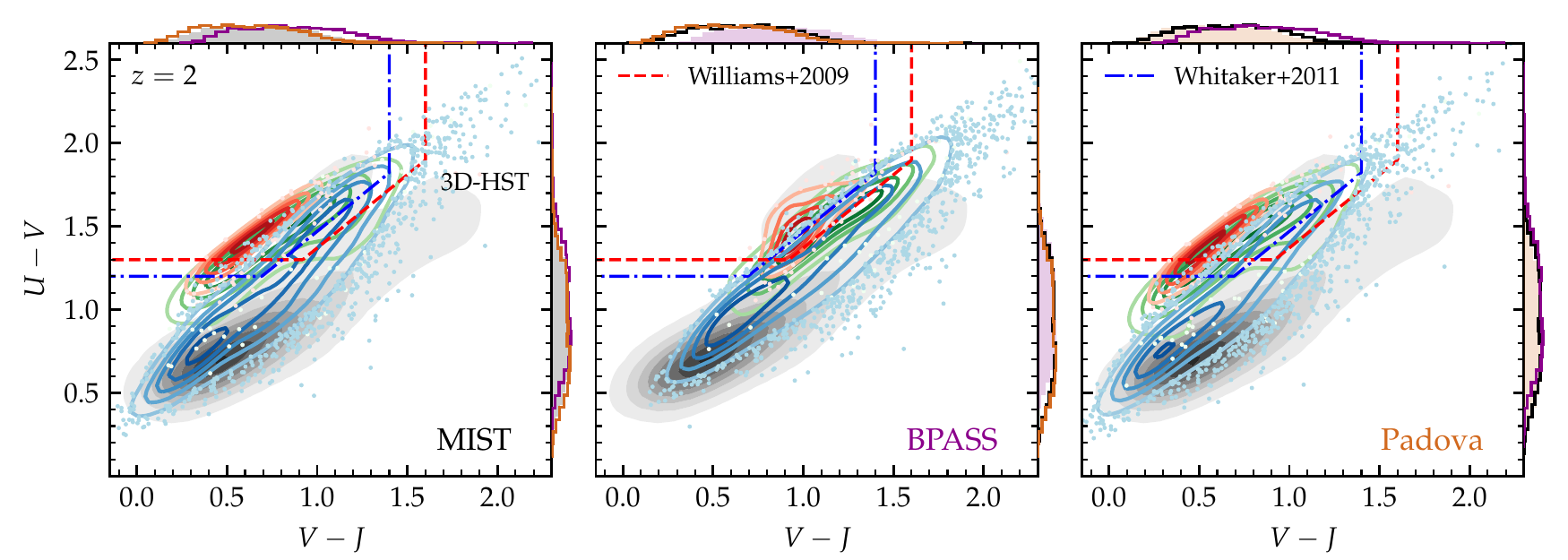}	
	\caption{Comparison of mock UVJ colors to 3D-HST observations. 
	Plot additionally shows the impact of different stellar isochrones on model UVJ diagram in \simba. Contours and observational UVJ selection criteria from \citet{williams_detection_2009} and \citet{whitaker_newfirm_2011} are shown as in Figure~\ref{fig:uvj_schematic}. We additionally show the outliers for each population as colored points. From left to right, the panels show UVJ diagrams computed from \pd\ using MIST, BPASS, and Padova isochrones.  Histograms on the axes show the distributions of $U-V$ and $V-J$ colors for each stellar isochrone choice, with the one particular to that panel filled in and the others as outlines.}\label{fig:uvj_isochrones} 
\end{figure*}

\subsubsection{Dependence on Stellar Population Models}\label{sec:uvj_obs_sps}

First, we examine how well \simba\ matches the observed distribution of rest-frame UVJ colors within the context of the underlying stellar model.
We do so by comparing the \simba+\pd\ UVJ diagram, under different model assumptions, to the observed sample from the 3D-HST survey at $1.7 < z < 2.3$ \citep{brammer_3dhst_2012, skelton_3dhst_2014, momcheva_3dhst_2016}.

Figure~\ref{fig:uvj_isochrones} shows the \simba\ UVJ diagram at $z=2$ for three different stellar population synthesis (SPS) models.
Specifically, we show the UVJ diagram for three different assumed stellar isochrones: MIST, which includes rotating stars \citep{choi_mesa_2016, dotter_mesa_2016, paxton_modules_2011}; BPASS, which includes binary stars \citep{eldridge_binary_2017}; and Padova \citep{bertelli_theoretical_1994, girardi_evolutionary_2000, marigo_evolution_2008}. 
While red, green, and blue contours show the distributions for quenched, transition, and star-forming galaxies in \simba, respectively, the gray contours show the 3D-HST sample. 
For each of the star-forming, transition, and quiescent populations, we plot the outliers as colored points.
On each panel, we show histograms of $U-V$ and $V-J$ colors for each isochrone.

\begin{figure}
	\centering
	\includegraphics[width=0.9\linewidth]{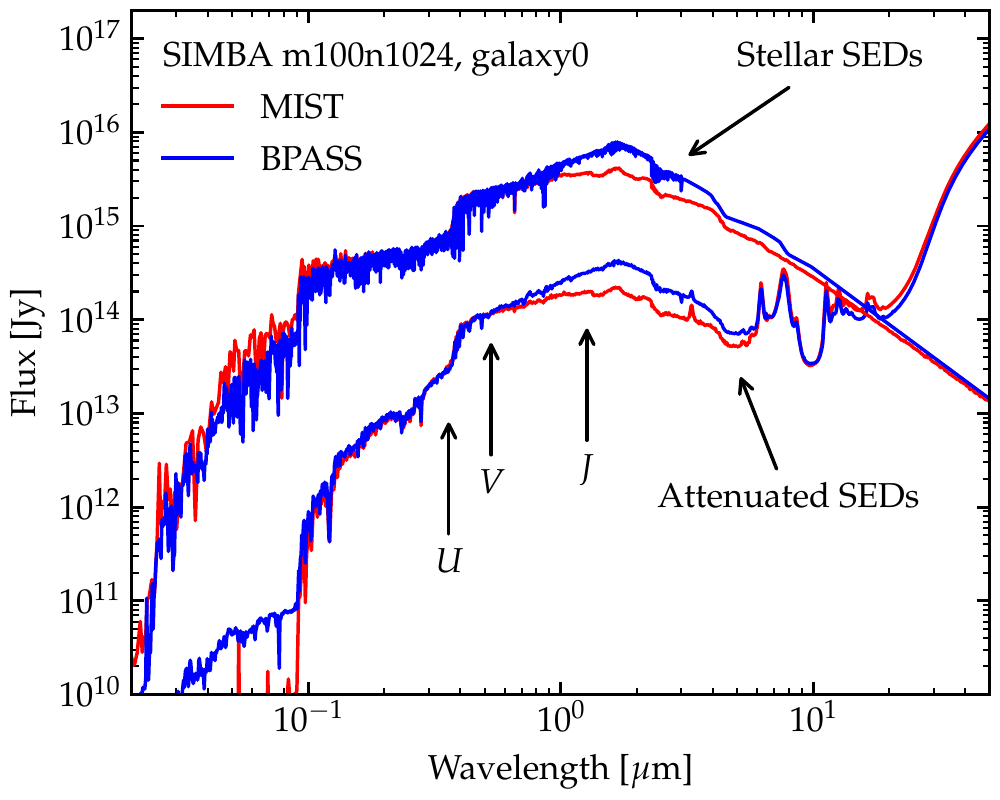}	
	\caption{SEDs for an example galaxy from our simulations  at $z=2$ using both MIST (red) and BPASS (blue) isochrones. Stellar and attenuated SEDs are shown, and we see that BPASS produces systematically redder colors due to a bump in the SED in the $J$-band.}\label{fig:bpass_SEDs}
\end{figure}

Immediately, it is apparent that there is some disagreement between the \simba\ UVJ diagram and the 3D-HST data. 
Specifically, while we predict a similar range of colors, we generally have bluer colors than the observational data. 
This is true for MIST and Padova isochrones; however, BPASS isochrones produce generally redder colors, more in line with the 3D-HST sample. 
Figure~\ref{fig:bpass_SEDs} shows (stellar and dust-attenuated) SEDs for an example galaxy computed with MIST and BPASS isochrones. 
We see that the systematically redder colors we get from BPASS isochrones are primarily due to a bump in the SED near the $J$ band. 

While the systematically redder colors produced from BPASS isochrones bring our results in closer agreement with observational selection criteria, they also noticeably change the distribution of galaxies on the UVJ diagram. 
The overlap between the quenched and star-forming populations is more significant in the BPASS model, with only a handful of quenched galaxies extending beyond the region occupied by SF galaxies. 
Additionally, BPASS places the oldest, reddest quiescent galaxies at the same $V-J$ as their younger counterparts, in contrast to observations \citep[e.g.][]{whitaker_quiescent_2013}.
As such, we adopt MIST isochrones in our fiducial SPS model, with the caveat that our colors are generally bluer than observations.
However, we interpret this as a systematic effect, and move forward with the assumption that our fiducial models broadly reproduce the observed UVJ diagram modulo this $\sim 0.2$--$0.3$ mag shift.
The significant contamination of the quiescent region from star-forming galaxies in our model is primarily due to this systematic color shift.

\begin{figure*}
	\centering
\includegraphics[width=0.87\linewidth]{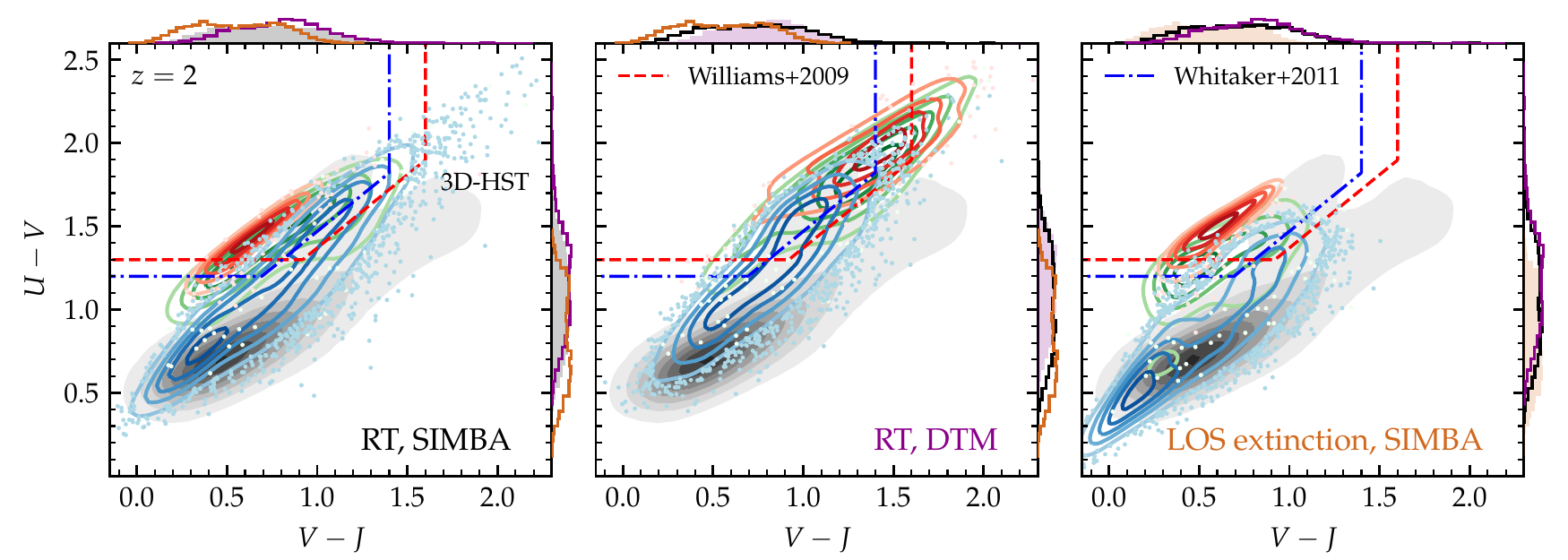}	
	\caption{Impacts of different dust models on the UVJ diagram in \simba. Contours and points are shown as in Figure~\ref{fig:uvj_isochrones}. From left to right, the panels show UVJ diagrams using \pd\ with the \simba\ explicit dust model, using \pd\ with a dust-to-metals radio of 0.4, and using a line-of-sight (LOS) extinction model with an sSFR and metallicity-dependent extinction law. Histograms on the axes show the distributions of $U-V$ and $V-J$ colors for each dust model choice, with the one of that panel filled in and others as outlines.}\label{fig:uvj_dustmodels}
\end{figure*}

\subsubsection{Dependence on Dust Models}\label{sec:uvj_obs_attenuation}

We next turn to understanding how the comparison between our model galaxies and those observed in UVJ space depends on the assumed underlying dust model.

Figure~\ref{fig:uvj_dustmodels} shows the \simba\ UVJ diagram at $z=2$ using three different dust models: the explicit dust model in \simba, in which gas particles keep track of dust creation, growth, and destruction on-the-fly in the cosmological simulation, a dust-to-metals model in which dust mass is assumed to scale with the metal mass by a constant ratio of 0.4, and a simplified line-of-sight (LOS) extinction model in which galaxies are assumed to follow an sSFR and metallicity-dependent extinction law. 
The former two models employ \pd\ radiative transfer in which dust is distributed throughout the galaxy as computed in the hydrodynamic galaxy formation simulations, whereas the latter model simply sums the LOS extinction from each star particle. 
This model is included primarily to demonstrate the importance of radiative transfer in computing realistic colors \citep[e.g][]{narayanan_powderday_2021}. 

It is clear from Figure~\ref{fig:uvj_dustmodels} that only the combination of the \simba\ explicit dust model and \pd\ radiative transfer is able to fully populate the dusty star-forming region of UVJ space (the top right of the diagram). 
These incredibly red, dusty galaxies have historically been a challenge for simulations to reproduce \citep[e.g.][]{dave_mufasa_2017, donnari_star_2019}, and even proved challenging for early observational surveys \citep[e.g.][]{williams_detection_2009} due to a lack of sufficiently red template SEDs in rest-frame color measurements \citep[see Appendix C in][]{whitaker_2010}.
As such, the success of our \simba+\pd\ dust model serves as an indicator of the critical importance of modeling dust physics explicitly when simulating broadband UVJ colors.
For example, the dust-to-metals model produces redder colors for quiescent galaxies than for dust star-forming galaxies, as these quiescent objects are some of the most metal-enriched objects in the simulation. 
This is inconsistent with the observed colors of galaxies at $z\sim 2$. 
At the same time, the LOS extinction model, even incorporating the \simba\ explicit dust masses, fails to properly populate the top right of the diagram.

Additionally, though not shown, we examine the effects of varying other model assumptions. 
We find little ($\sim 0.03$ mag) difference in UVJ colors resulting from varying the stellar initial mass function (IMF) between those of \citet{kroupa_initial_2002}, \citet{chabrier_galactic_2003}, and \citet{salpeter_luminosity_1955}. 
We find a comparably small difference in UVJ colors from varying the spectral library from the MILES \citep{sanchez-blazquez_mediumresolution_2006} and BaSeL \citep{westera_standard_2002} libraries. 
We find nearly no variation in UVJ colors from the inclusion of attenuation by circumstellar AGB dust \citep[using the model of][]{villaume_circums_2015}, AGN emission and dust model using SED templates from \citet{nenkova_agn_2008, nenkova_agn_2008a}, or nebular line and continuum emission from \textsc{cloudy} lookup tables \citep{narayanan_powderday_2021, byler_nebular_2017, byler_stellar_2018, byler_selfconsistent_2019, Garg_2022}.

\subsection{Galaxy Physical Properties in UVJ Space}\label{sec:uvj_trends}

We have demonstrated that \simba+\pd\ can broadly reproduce the observed distribution of UVJ colors, with a few notable exceptions: we do not reproduce the clear bimodal number density in UVJ space, and we produce systematically bluer colors than observations.
To further interrogate these inconsistencies with observations, we investigate the distribution of galaxy properties on the UVJ diagram and compare, qualitatively, to observed trends.

\subsubsection{Dust Attenuation}\label{sec:uvj_av}

\begin{figure}
    \centering
    \includegraphics[width=\linewidth]{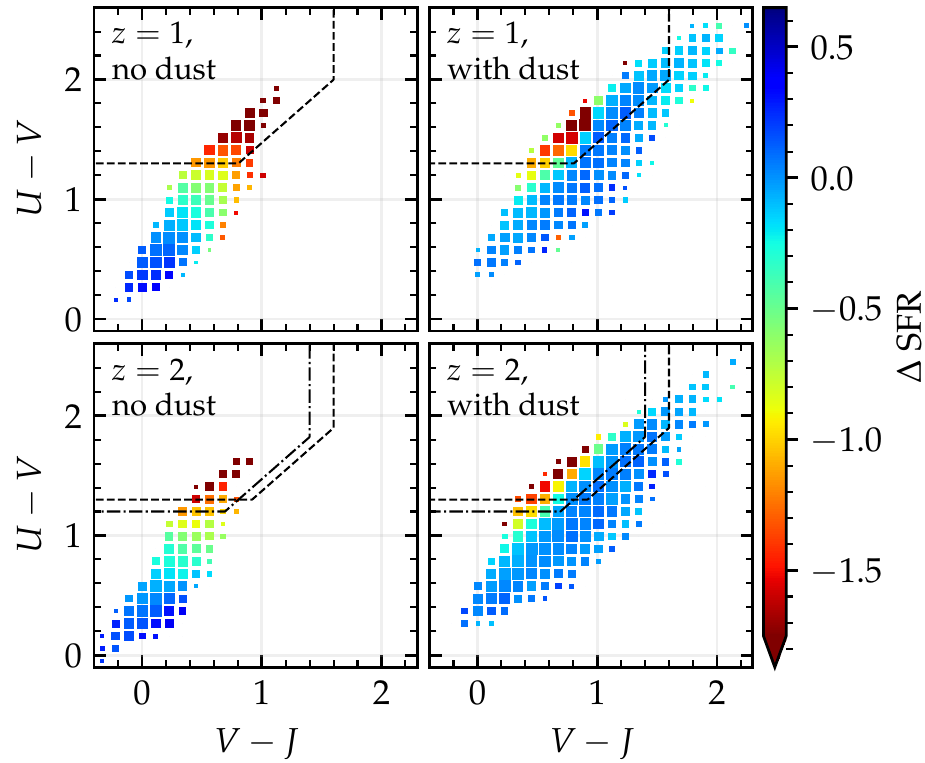}
    \caption{The \simba\ UVJ diagram at $z=1$ (top) and $z=2$ (bottom) both including the effects of dust (right column) and ignoring all dust (left column). We show the median $\Delta$SFR in bins of $U-V$ and $V-J$, with points sized logarithmically according to the number of galaxies in the bin. While our dust-free models produce a clear gradient in $\Delta$SFR according to $U-V$ color, including the effects of dust in our radiative transfer models moves star-forming galaxies to redder colors.}
    \label{fig:dust_content}
\end{figure}

We begin with an examination of trends in the dust attenuation $A_V$ in our model UVJ diagram at $z=1-2$.
In Figure~\ref{fig:dust_content}, we show our mock UVJ colors for our model galaxies at $z=1$ and $z=2$ for both a model without dust (i.e. discarding all of the dust content in our radiative transfer models) as well as a model including our fiducial dust model.
Generally, without dust, galaxies form a relatively tight locus with a clear trend in $\Delta$SFR.
This is in line with dust-corrected UVJ diagrams inferred from observations \citep[e.g.][]{fang_demographics_2018}.  
Including dust decreases the fidelity of this trend significantly as dust reddening pushes star-forming galaxies toward redder $V-J$ colors.
We note that there are a small number of quiescent galaxies at $z=1$ with dust-free colors placing them outside the quiescent region. 
This owes to a recent frosting of star-formation, which we explore further in Section~\ref{sec:uvj_age}.

\begin{figure*}
	\centering
	\includegraphics[width=\linewidth]{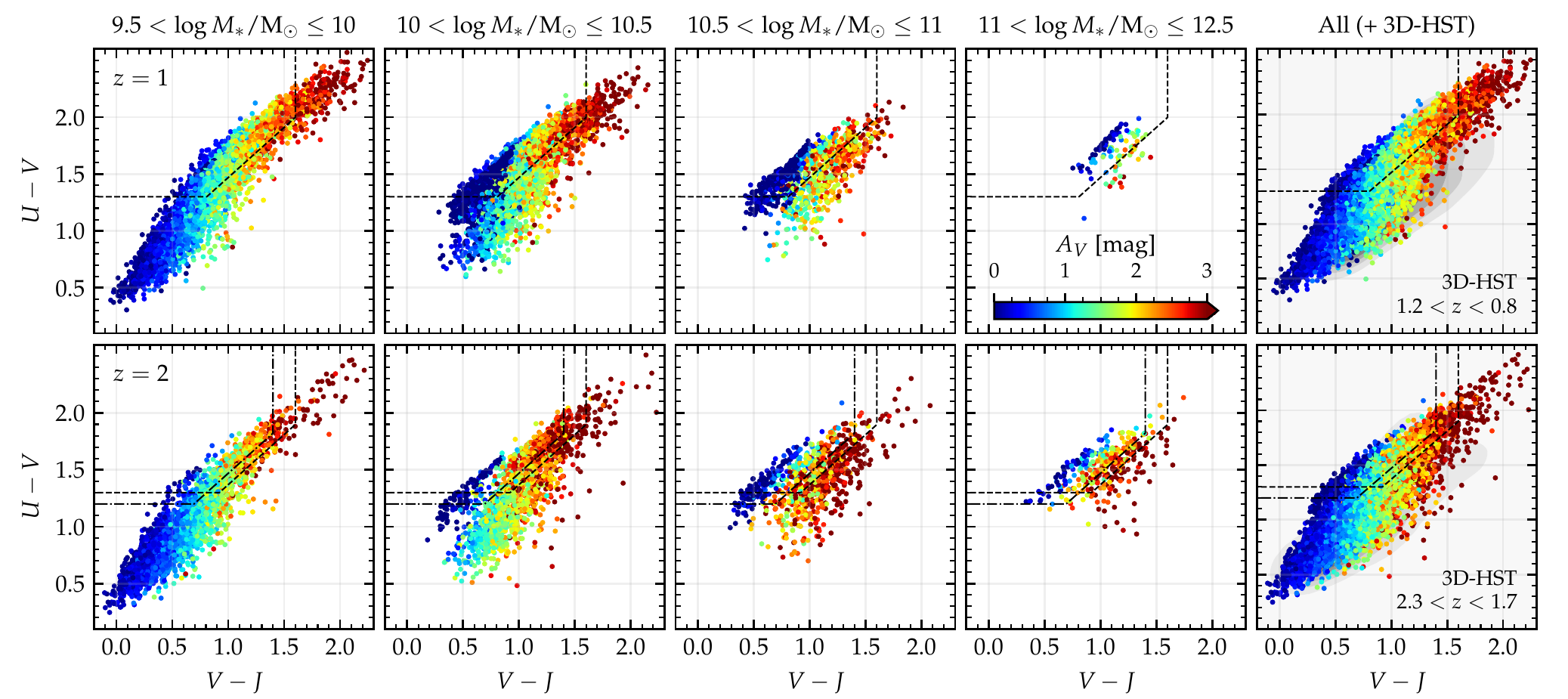}
	\caption{The \simba\ UVJ diagram in 4 bins of stellar mass. 
	The top (bottom) row shows data for $z=1$ ($z=2$), and points are colored by the $V$-band dust attenuation $A_V$. The right column shows all model galaxies overplotted with KDE contours showing observations from the 3D-HST survey \citep{brammer_3dhst_2012, momcheva_3dhst_2016}. 
	A strong correlation between $A_V$ and $V-J$ color is evident at low masses, but breaks down at higher masses, where many heavily attenuated galaxies have low $V-J$.}\label{fig:uvj_A_V}
\end{figure*}

In Figure~\ref{fig:uvj_A_V}, we show our galaxies in UVJ space, color-coded by $A_{V}$.
We first highlight the right column, which shows the entire model galaxy sample for $z \sim 1$ (top), and $z \sim 2$ (bottom), with contours showing observations from the 3D-HST survey \citep{brammer_3dhst_2012, skelton_3dhst_2014, momcheva_3dhst_2016}. 
Consistent with observational constraints \citep[e.g.][]{price_direct_2014, martis_evolution_2016, fang_demographics_2018}, we see lines of constant $A_V$ for star-forming galaxies at roughly constant $V-J$.
This said, this is a nuanced and mass-dependent trend. 
While we see a strong correlation between $A_V$ and $V-J$ color for star-forming galaxies in the $9.5 < \log M_*/\Msun \leq 10$ mass bin, at $10 < \log M_*/\Msun \leq 10.5$ several highly dusty ($A_V \gtrsim 2.5$) galaxies show bluer $V-J$ colors than expected, and this trend continues to weaken at higher masses.
Observational surveys \citep[e.g.][]{fang_demographics_2018} tend to find high-mass ($\log M_*/\Msun > 10$) galaxies in the DSFG region on the far top right of the UVJ diagram, and low-mass ($\log M_*/\Msun < 10$) galaxies concentrated in the bottom left.
The population of low-mass, highly dust-reddened galaxies in \simba\ is likely a byproduct of our dust evolution model \citep{li_dusttogas_2019}, as observations typically do not find galaxies at this mass with $A_V\gtrsim 1$.
Thus, while we successfully populate the dusty star-forming region of the UVJ diagram, we do so primarily with low-mass ($\log M_*/M_\odot \lesssim 10.5$) galaxies, rather than higher-mass galaxies as is seen in observational surveys. 
This, alongside the lack of a clear bimodality in our UVJ diagram, represent notable tensions with observations.

\begin{figure*}
    \centering
    \includegraphics[width=0.7\linewidth]{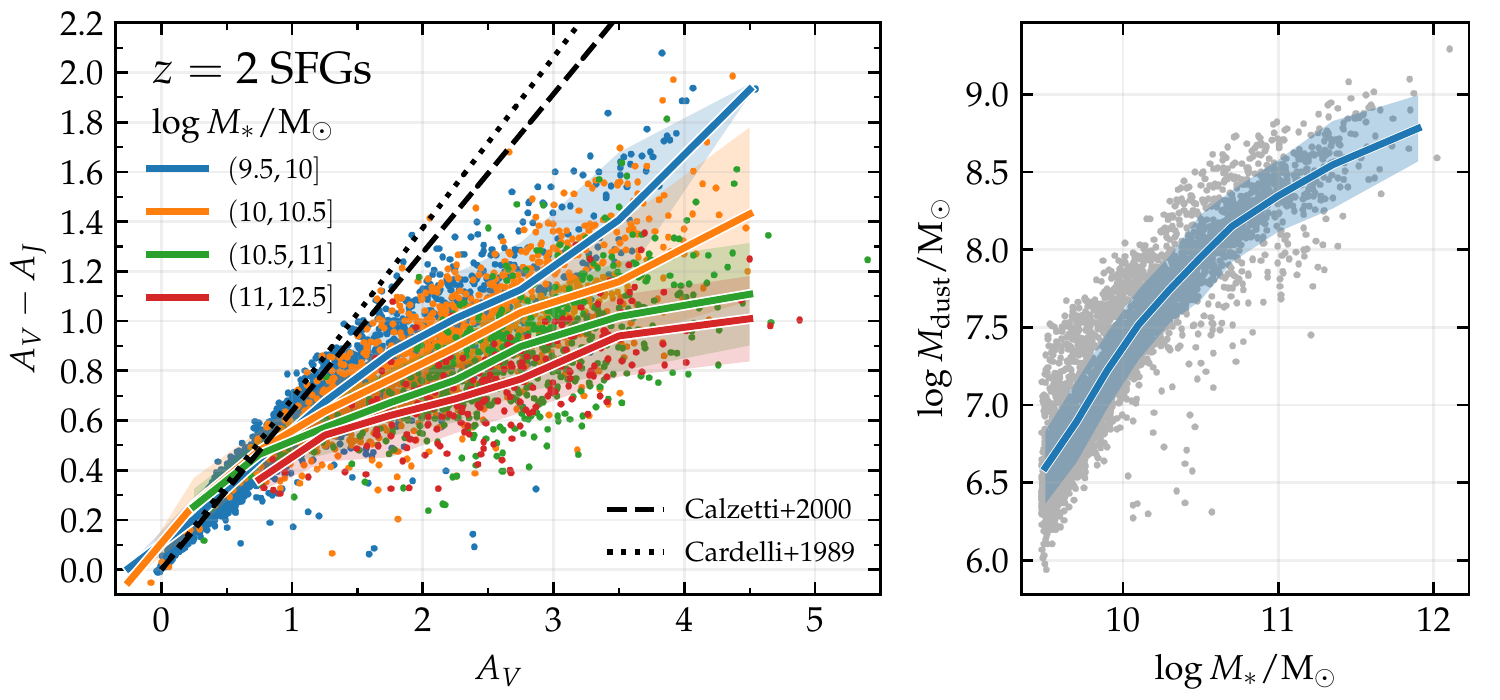}
    \caption{{\it Left:} Relationship between dust reddening ($A_V-A_J$) and $A_V$ for star-forming galaxies at $z=2$. 
    We bin the sample into four bins of stellar mass (as in Figure~\ref{fig:uvj_A_V}) and plot (for each mass range) the median $A_V-A_J$ in 10 bins of $A_V$.
    We see that, at the same $A_V$, higher-mass galaxies experience less dust-reddening.
    {\it Right:} Scatter plot of dust mass vs.~stellar mass for the same sample of star-forming galaxies at $z=2$, with KDE contours overlaid. 
    We see that our high-mass galaxies still retain high dust masses, suggesting that the lack of high-mass galaxies in the dusty star-forming region of UVJ space is driven by grayer attenuation curves rather than a lack of dust content. }
    \label{fig:dust_reddening}
\end{figure*}

To further explore the relationship between mass, $A_V$, and attenuation curve slope (which impacts galaxy location in UVJ space), we show in Figure~\ref{fig:dust_reddening} the amount of dust-reddening (defined as $A_V - A_J$) as a function of $A_V$ for star-forming galaxies in \simba\ at $z=2$.
$A_V-A_J$ effectively measures the slope of the attenuation curve between the $V$ and $J$ bands, and galaxies with higher $A_V-A_J$ will be found further to the right in UVJ space. 
We bin the sample into four bins of stellar mass and plot the median $A_V-A_J$ in each bin.
We additionally plot the relationship between $A_V-A_J$ vs $A_V$ for the literature attenuation curves of \citet{cardelli_relationship_1989} and \citet{calzetti_dust_2000}.
We see that at $A_V \gtrsim 1$, our model galaxies generally show grayer attenuation curves than is expected for galaxies at this redshift.
Moreover, at the same $A_V$, higher-mass galaxies in \simba\ experience less dust-reddening, i.e. they have grayer attenuation curves. 
That is, as our simulated galaxies become more massive, the star-dust geometry becomes increasingly complex, and the attenuation curves become flatter (grayer) as stars and dust are spatially decoupled.
This results in less pronounced reddening for the most massive galaxies.

As a check, we also show in the right panel of Figure~\ref{fig:dust_reddening} a scatter plot of the dust masses vs. the stellar masses of the same sample of $z=2$ SFGs.\footnote{While we show these trends only for $z=2$ star-forming galaxies, we have confirmed that they hold at $z=1$. We show these trends only for star-forming galaxies, as quiescent galaxies in \simba\ almost universally have $A_V \sim 0$.}
We see that our high-mass galaxies still retain significant dust masses.
This is consistent with recent observational constraints by \citet{shapley_first_2020} and \citet{dudzeviciute_tracing_2021}, which have found that the \simba\ dust model reasonably reproduces the dust to gas ratio and dust mass function at $z\sim 2$. 
Similarly, the recent review by \citet{peroux_cosmic_2020} shows that the \simba\ model accurately reproduces observational constraints on the evolution of the cosmic dust density.
This suggests that the absolute dust contents in our model galaxies are reasonable, and reaffirms that the lack of high-mass, highly reddened galaxies must be due to the lack of obscuration and relatively gray attenuation laws.

Therefore, one possibility for reducing tensions with observations would be if our galaxies had steeper attenuation laws. 
Indeed, some observations have inferred laws steeper than those presented in our models in Figure~\ref{fig:dust_reddening} \citep[e.g.][]{mclure2018}.  
While a quantitative comparison between the attenuation curves for star-forming galaxies in our model and those observed are outside the scope of this paper \citep[though see][]{salim_dust_2020a}, we note that the observational derivation of dust attenuation curves from unresolved systems at high-$z$ comes with significant attendant uncertainties, including assumed SED shapes (and location in IRX-$\beta$ space), as well as the shape of the intrinsic stellar continuum \citep{narayanan_irxb_2018,reddy2018}.
Therefore, we move on with the assumption that our dust model reasonably reproduces the observed UVJ diagram, though fails to produce adequate dust reddening in the highest-mass galaxies.

\subsubsection{Star Formation Rates} \label{sec:uvj_sfr}

\begin{figure*}
	\centering
	\includegraphics[width=\linewidth]{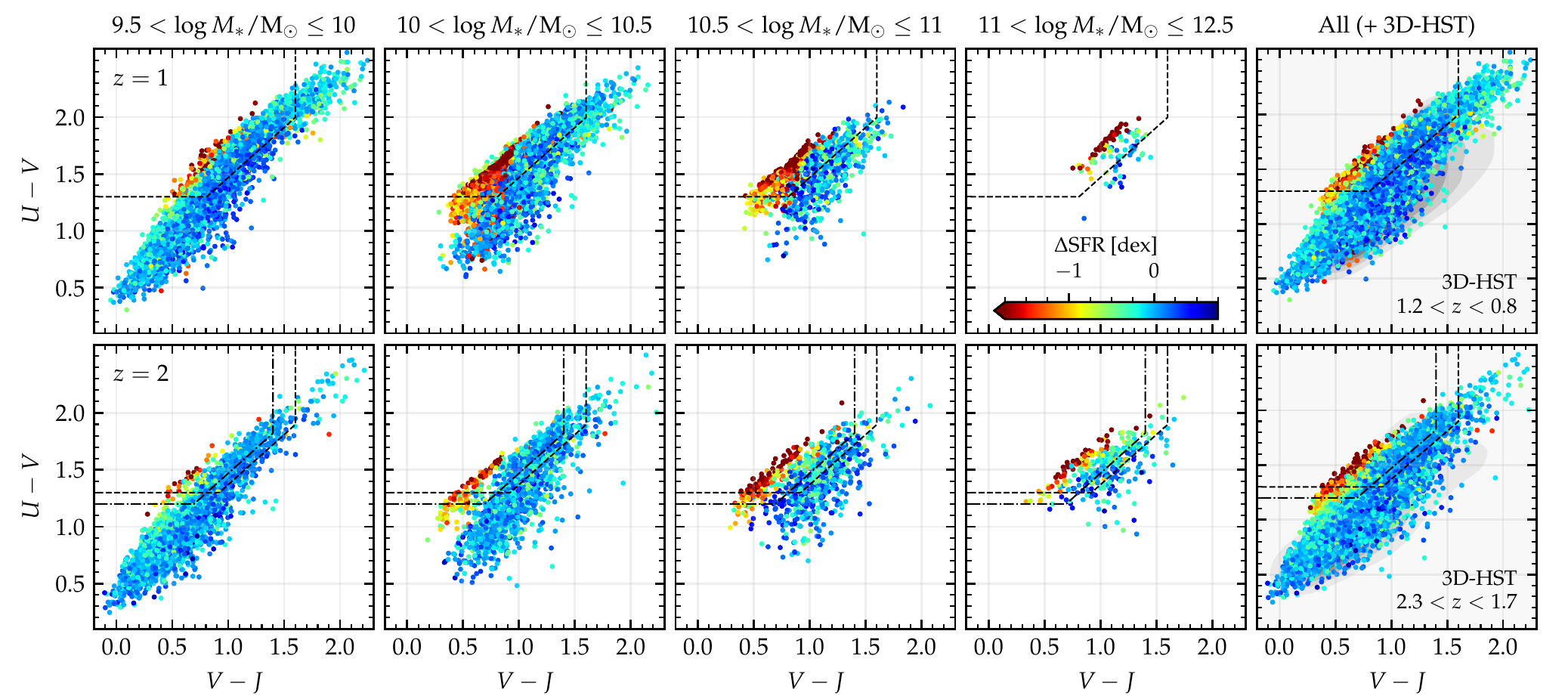}
	\caption{Same as Figure~\ref{fig:uvj_A_V}, but with points colored by the distance from the star-forming main-sequence, $\Delta$SFR. 
	We see that star-forming and quiescent galaxies are well separated by the UVJ selection line, and a mass trend is evident in the star-forming population.}\label{fig:uvj_sfr}
\end{figure*}

We now examine trends with galaxy star formation rates in UVJ space.
In Figure~\ref{fig:uvj_sfr}, we show the $z=1$ and $z=2$ UVJ diagrams in 4 bins of stellar mass.
Points are colored by each galaxy's distance, in dex, from the SFMS ($\Delta$SFR). 
Despite the lack of a clear bimodality, quiescent and star-forming galaxies are well separated in UVJ space at all mass ranges, with quiescent galaxies occupying a narrow locus on the top left and star-forming galaxies populating the blue cloud.
In between these two populations lie transition galaxies.

\begin{figure*}
	\centering
	\includegraphics[width=\linewidth]{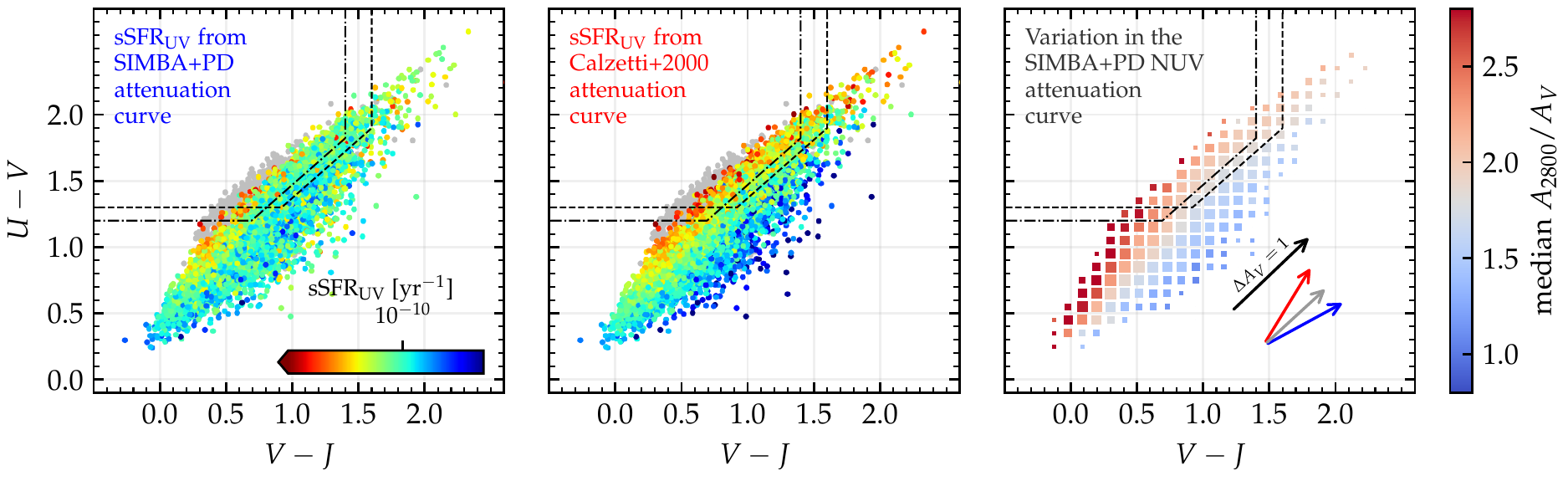}
	\caption{Star-forming galaxies on the UVJ diagram at $z=2$, with points colored by dust-corrected UV-derived sSFRs. On the left panel, dust correction is done using the true \simba$+$\pd\ attenuation curve. In the middle panel, dust correction is done assuming a universal \citet{calzetti_dust_2000} attenuation curve as described in the text. The right panel shows the median ratio of the NUV attenuation $A_{2800}$ to $A_V$, in bins of $U-V$ and $V-J$. The colorbar is centered on $A_{2800}/A_V = 1.8$, the value for the Calzetti curve. Points are sized logarithmically according to the number of galaxies in that bin. Arrows show the ``dust vector,'' or the effect on the UVJ colors of adding $\Delta A_V = 1$, for the Calzetti curve (black) as well as the median and the 25th and 75th percentile \simba\ curves (red, gray, and blue). Due in large part to varying dust vectors, the optical-NUV slope of the attenuation curve is correlated with the UVJ colors of star-forming galaxies. As such, the assumption of a universal attenuation curve exaggerates the subtle trend in sSFR on the UVJ diagram.}\label{fig:uvj_sfr_calzetti}
\end{figure*}

Numerous observational studies have found ``stripes'' of constant sSFR running roughly parallel to the diagonal selection line in UVJ space, with the youngest, most actively star-forming galaxies along the bottom right \citep{williams_detection_2009, williams_evolving_2010, patel_starformationratedensity_2011, whitaker_large_2012, fang_demographics_2018, leja_uvj_2019}.
This trend has generally been interpreted as support for the efficacy of UVJ selection of quiescent galaxies.
This sSFR trend is somewhat subtle in our simulations, and at face value, appears at odds with observational constraints. 
We posit that tension is simply a manifestation of the diverse nature of dust attenuation curves in high-$z$ galaxies.
When deriving properties from SEDs at high-redshift, many observations assume a universal \citet{calzetti_dust_2000} dust attenuation law; in contrast, our simulations \citep[as do many, e.g.][]{narayanan_theory_2018,trayford_fade_2020,lagos_physical_2020} yield wildly varying attenuation curves for high-$z$ galaxies. 
This mismatch in assumed versus actual attenuation curve can bias properties derived from dust-corrected SEDs.

To examine this, we calculate dust-corrected UV SFRs following \citet{fang_demographics_2018} as 
\begin{equation}
  \mathrm{SFR}_{\mathrm{UV}}~[\mathrm{M}_\sun~\mathrm{yr}^{-1}]=2.59\times 10^{-10}~L_{2800} ~ 10^{0.4\, A_{2800}}
\end{equation}
where $L_{2800}$ is the $2800$~\AA~luminosity from the SED (in $\mathrm{L}_\sun$) and $A_{2800}$ is the corresponding attenuation (in mag). 
We perform these calculations in two manners: first by using the actual $A_{2800}$ values from \pd\ (i.e. employing the true dust attenuation curve for our model galaxies) and second using the assumption of a \citet{calzetti_dust_2000} law where $A_{2800} = 1.8 A_V$. 
Figure~\ref{fig:uvj_sfr_calzetti} shows how these different attenuation curve assumptions produce different estimates of the sSFR. 
We show star-forming galaxies at $z=2$, with points colored by $\mathrm{sSFR}_{\mathrm{UV}}$ computed assuming \pd\ attenuation curves (left panel) and assuming a Calzetti curve (middle panel).

Different attenuation curve assumptions---either a universal or widely varying attenuation curve---can impact trends in sSFR in UVJ space.
To illustrate this, we show in the right panel of Figure~\ref{fig:uvj_sfr_calzetti} the relationship between galaxy positions in UVJ space and the optical-NUV slope of the \simba$+$\pd\ attenuation curve. 
The colormap in this panel is centered on $A_{2800}/A_V = 1.8$, the value for a Calzetti law. 
We see that, for star-forming galaxies, the shape of the attenuation curve is correlated with the galaxy's location on the UVJ diagram. 
In particular, galaxies near the quenched region have a steeper attenuation curve in the $U-V$ and therefore have a steeper ``dust vector'' (the arrows shown in Figure~\ref{fig:uvj_sfr_calzetti}). 
For galaxies on the bottom right, the opposite is true.
This implies that, if the true attenuation curves indeed vary, the assumption of a universal Calzetti law would tend to underestimate SFRs near the quenched region and overestimate along the bottom right.

There is evidence in the literature that the spread of star-forming galaxies in UVJ space is driven by structural properties and observed inclination \citep{patel_uvj_2012}, and this has been supported by semi-analytical models \citep{zuckerman_reproducing_2021}.
Such variations in the star-dust geometry will lead to varying attenuation curves \citep{narayanan_theory_2018}, and assuming a universal attenuation curve in spite of this variation will produce a stronger observational correlation between SFR and location in UVJ space.

In summary, while the trends we observe in sSFR on the UVJ diagram are more subtle than observed trends, this can be largely attributed to the fact that observations tend to assume a universal attenuation curve in spite of underlying variation.
A full analysis of the dependence galaxy properties derived from SEDs on the assumed attenuation curve is beyond the scope of this paper \citep[though is explored in more detail in][]{lower_how_2020a,lower_how_2022}.
This said, we note that a continual trend in sSFR on the UVJ diagram has been reproduced even with SED fitting codes that allow for a varying attenuation curve \citep[e.g.][]{leja_uvj_2019}.

\subsubsection{Stellar Age}\label{sec:uvj_age}

\begin{figure}
	\centering
	\includegraphics[width=\linewidth]{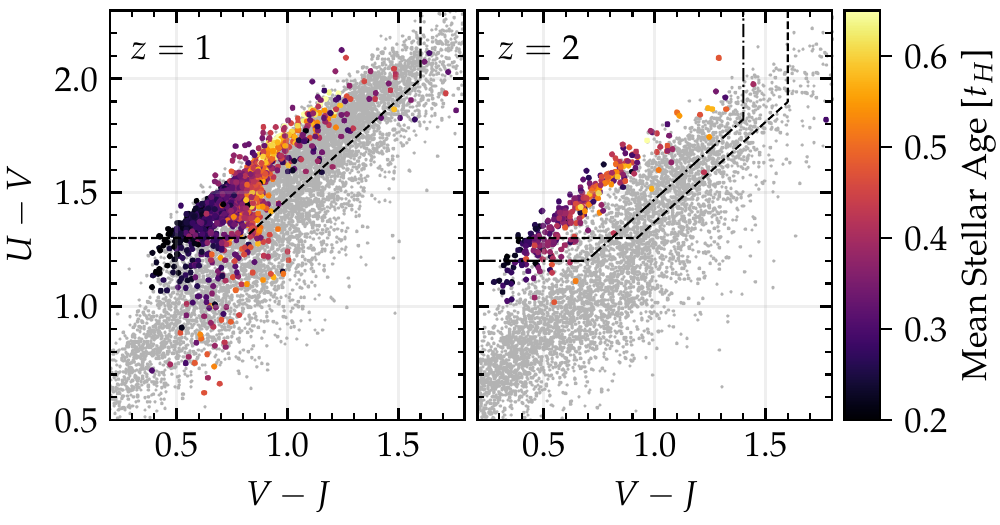}	
	\caption{Gradient in stellar age for quiescent galaxies on the UVJ diagram. The left (right) panel shows the UVJ diagram at $z=1$ ($z=2$), and points are colored by the mass-weighted mean stellar age.}\label{fig:uvj_age}
\end{figure}

Finally, we examine trends in stellar age on the UVJ diagram. 
Figure~\ref{fig:uvj_age} shows quiescent galaxies on the UVJ diagram at $z=1$ and $z=2$, with points colored by the mass-weighed mean stellar age as a fraction of the Hubble time.
We compute mass-weighted mean stellar ages by averaging the formation times of star particles, weighted by the formation masses, computed using an \textsc{fsps} Simple Stellar Population to account for mass-loss by evolved stars.
We show only quiescent galaxies in Figure~\ref{fig:uvj_age} as we do not observe a trend in stellar age for star-forming galaxies. 
Though this is in contrast to the observations of \citet{whitaker_large_2012}, it is consistent with the subtlety of the trend we see with sSFR in UVJ space. 

A clear trend in mean stellar age has been observed in the quiescent region of UVJ space \citep{whitaker_large_2012,whitaker_quiescent_2013,belli_mosfire_2019}.
\citet{leja_uvj_2019} showed that this trend in stellar age, along with trends in metallicity, is not perfectly constrained by UVJ colors alone, and instead is a result of more fundamental galaxy scaling relationships.
Regardless, as an oft-used observable, it is a fruitful comparison to investigate trends with stellar age in UVJ space.
We reproduce the observed gradient in stellar age reasonably well, with the oldest galaxies universally occupying the top right of the quiescent population. 
Furthermore, we find that the youngest quenched galaxies typically lie in the lower left of the quenched region, consistent with observations of post-starburst (PSB or E$+$A) galaxies \citep[e.g.][]{yano_relation_2016, almaini_massive_2017, suess_color_2020}.
The age gradient in the quenched region implies a fairly universal, predictable evolution of galaxies on the UVJ diagram once they quench. 

However, the inferred evolution of quiescent galaxies in UVJ space is complicated by the nontrivial portion of quenched galaxies that lie outside the quenched region, particularly at $z=1$. 
These aberrant galaxies show slightly higher specific star-formation rates than the rest of the quenched population, but are generally older than we would expect based on a simple age gradient along the diagonal in Figure~\ref{fig:uvj_age}. 
These galaxies have bluer $U-V$ colors than galaxies in the quenched region owing to a recent frosting of star formation \citep[e.g.][]{ford_direct_2013,haines_constructing_2013,akhshik_recent_2021}. 
Indeed, $\sim 60\%$ of these galaxies have a ratio of their averaged star formation rates SFR$_{50}$/SFR$_{200} > 1$, compared to $\lesssim 10\%$ for quenched galaxies in the quenched region.

\section{Time Evolution in UVJ Space}\label{sec:evolution}

\begin{figure*} 
	\centering
	\includegraphics[width=\linewidth]{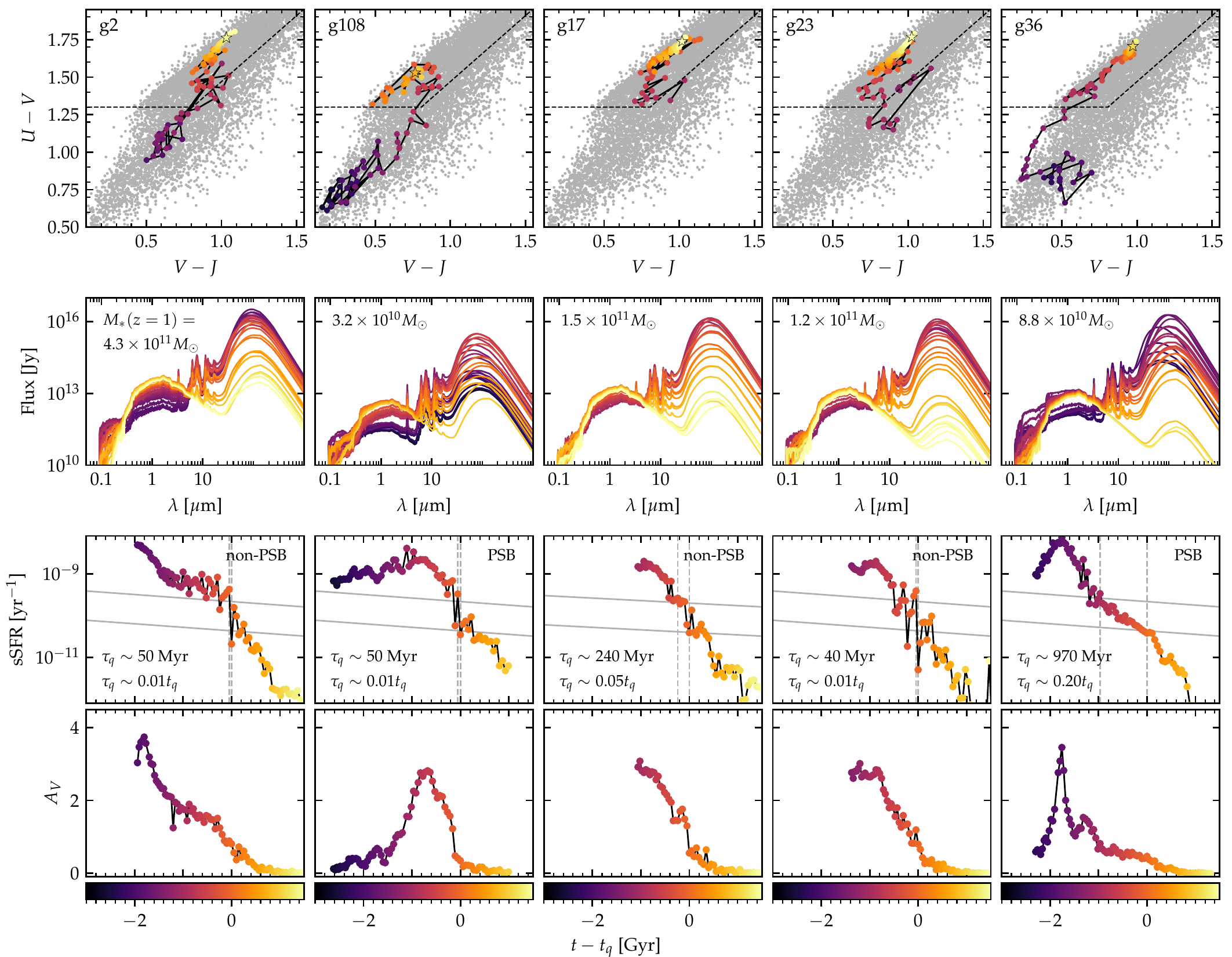}
	\caption{Evolution of UVJ colors from $z\sim 2.6-1$ for 5 galaxies in the process of quenching. In all panels, points are colored by time since quenching. The top row shows $U-V$ vs.~$V-J$ colors for each galaxy, and the stars indicate the galaxy locations at $z=1$. The next row shows the SEDs at every third snapshot, with the $z=1$ stellar mass annotated. The third row shows the sSFR, with solid lines indicating the sSFR thresholds for used to define quenching times and dashed lines indicating the start and end times of quenching. We additionally categorize the galaxies as ``PSB'' or ``non-PSB'' as described in Section~\ref{sec:psbs}. The bottom row shows the dust attenuation $A_V$. Open circles on the time series panels indicate where galaxies are within the quenched region of UVJ space.}\label{fig:time_ev_examples}
\end{figure*}

We have shown that the \simba\ and \pd\ models broadly reproduce the observed UVJ diagram, and we have explored the factors driving inconsistencies with observations.
With some confidence that our simulations reasonably reproduce observations, we now turn our attention to understanding the physics that drives the evolution of galaxies in UVJ space.
Though much of the analysis presented thus far has focused on the flagship $100~\mathrm{Mpc}~h^{-1}$ \simba\ run, we now turn our attention to higher-resolution $25~\mathrm{Mpc}~h^{-1}$ \simba\ run, which outputs twice as many snapshots and thus provides substantially improved time resolution.  
Because of the smaller box-size---and hence, higher mass resolution---the $25~\mathrm{Mpc}~h^{-1}$ box has significantly more low-mass galaxies identified than our fiducial $100~\mathrm{Mpc}~h^{-1}$ box.
Therefore, in order to compare our results directly to the analysis presented thus far, we select only those galaxies with $\log M_*/\Msun > 9.5$ at $z=1$.
We further limit our analysis to only those massive galaxies that are quenched by $z=1$. 
We trace progenitors of these galaxies from $z\sim 2.6$ to $z=1$ in order to study their evolution as they quench. 
Of the 24 galaxies in our sample of massive, quenched galaxies at $z=1$, we find that 4 galaxies were quenched before $z\sim 2.6$ and thus do not experience a ``quenching event'' in the timespan tracked.
As our goal is to explore how galaxies evolve in color-color space as they quench, we do not include these galaxies in the subsequent analysis.  

Figure~\ref{fig:time_ev_examples} shows the evolution of $U-V$ and $V-J$ colors, SEDs, sSFR, and $A_V$ for five of the galaxies tracked.
In all panels, points are colored by $t - t_q$, the time since quenching. 
We define $t_q$ as the time at which a galaxy first drops below ${\rm sSFR} = 0.2~t_H^{-1}$, where $t_H$ is the age of the universe at that epoch \citep[following][]{pacifici_evolution_2016, rodriguezmontero_mergers_2019}.
The top row shows the UVJ diagram and the second row shows attenuated SEDs at every third snapshot.
The third row shows star formation histories, with solid lines indicating the relevant sSFR thresholds and dashed lines indicating the start and end times of quenching. 
The bottom row shows $A_V$ as a function of time.  
We refer to these galaxies by their IDs, listed in the corners of the top panel in each column in Figure~\ref{fig:time_ev_examples}. 
We show the evolution of the UVJ colors for the remaining 16 galaxies in Figure~\ref{fig:time_ev_examples_all}.

From Figure~\ref{fig:time_ev_examples}, it is evident that there are a diversity of quenching pathways through the UVJ diagram. 
Some galaxies, like galaxy 2, consistently move towards the quenched region as they evolve and enter the quenched region along the diagonal section of the boundary. 
Others, like galaxies 6 and 17, evolve more chaotically through the blue cloud but enter the quenched region in a similar fashion. 
Others still enter the quenched region from the bottom left. 
Galaxy 108 moves upwards and to the right on the UVJ diagram as it gets dustier, and then moves to the left and enters the quenched region near the bottom left. 
Galaxy 36, in contrast, moves rapidly to the left edge of the blue cloud as it forms a burst of stars and declines in dust attenuation, entering the quenched region from the bottom left. 

The diversity of quenching pathways in UVJ space is consistent with the diversity of SFHs we find in \simba\ and the dependence of dust attenuation on the dust geometry of the galaxy. 
We now further investigate the tracks that galaxies take in UVJ space, investigating the dependencies on quenching timescale and SFH.

\begin{figure*}
	\centering
	\includegraphics[width=\linewidth]{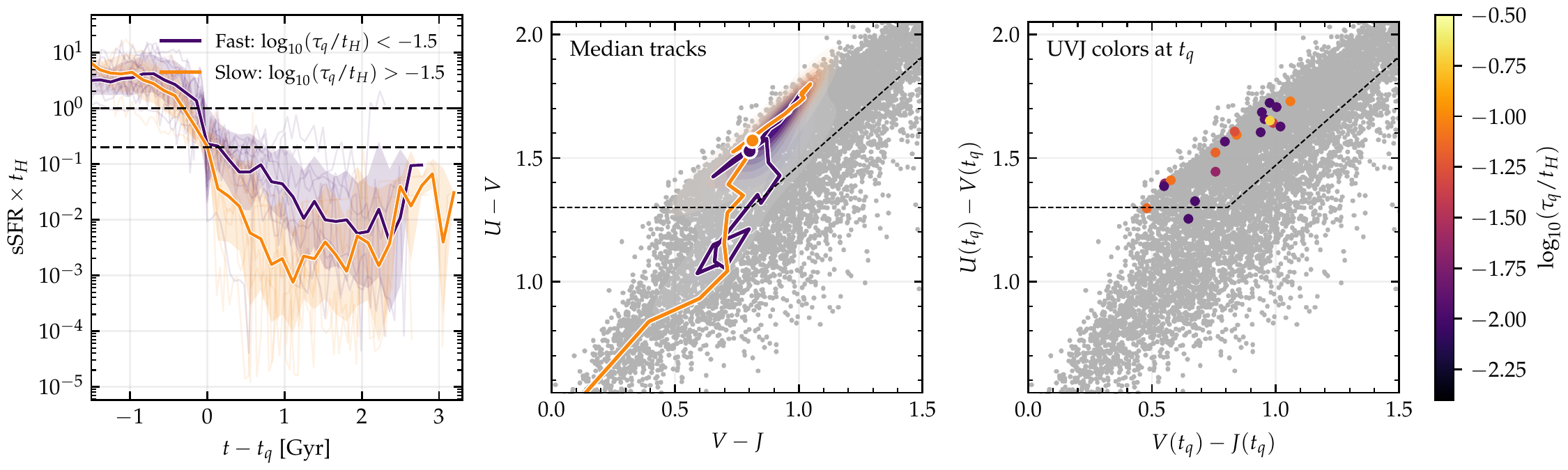}
	\caption{Evolutionary tracks in UVJ space as a function of the quenching timescale $\tau_q$. \textit{Left:} Median SFHs for fast and slow-quenching galaxies. The sSFR is scaled by $t_H$ to match the definition of quenching timescale, and the $x$-axis is centered on the time of quenching $t_q$. \textit{Middle:} Median evolutionary tracks in UVJ space, in bins of $t-t_q$, for fast and slow-quenching galaxies. Points indicate the median UVJ colors at $t=t_q$. We show Kernel Density Esimate (KDE) contours of the distribution of UVJ colors for all fast and slow-quenching galaxies at all timesteps in order to highlight the diversity underlying the median. \textit{Right:} UVJ colors at the time of quenching, colored by the quenching timescale. We see no clear evidence for different evolutionary tracks for fast- vs. slow-quenching galaxies.}\label{fig:time_ev_tau_q} 
\end{figure*}

\begin{figure*}
	\centering
	\includegraphics[width=\linewidth]{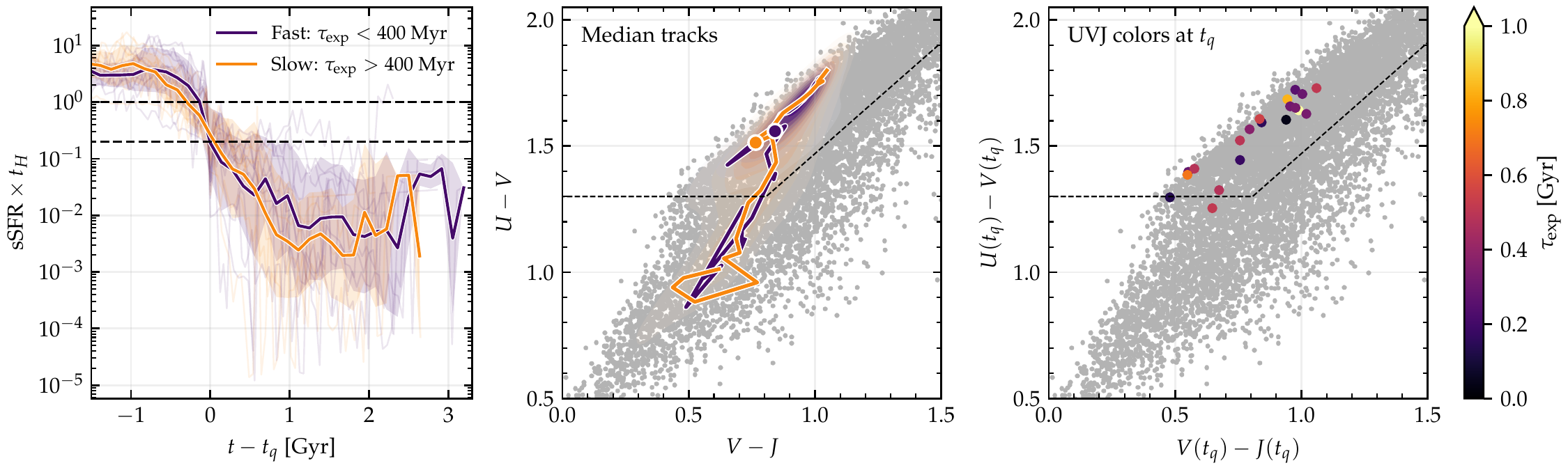}
	\caption{Evolutionary tracks in UVJ space as a function of the SFH decay timescale $\tau_{\rm exp}$. We again see no clear evidence for different evolutionary tracks for fast- vs. slow-quenching galaxies. }	
	\label{fig:time_ev_tau_exp}
\end{figure*}

\subsection{Fast vs.~slow quenching}

First, we explore the dependence of color-color evolution on the quenching timescale. 
Recent observations suggest that galaxies that quench on different timescales may trace different paths. 
In particular, \citet{belli_mosfire_2019} explore toy models for the UVJ colors of fast-quenching and slow-quenching galaxies at $1.5 < z < 2.5$.
They find that a fast-quenching galaxy (a tau-model SFH with a decay timescale $\tau \sim 100$ Myr) would typically enter the quenched region of UVJ space from the bottom left, while a slow-quenching galaxy (with $\tau \sim 1$ Gyr) would typically enter along the diagonal line, at redder colors.
Similarly, \citet{carnall_vandels_2019} explore the UVJ evolution of massive quiescent and green valley galaxies in the VANDELS survey at $1.0 < z < 1.3$. 
They find a typical model track that enters the quiescent region along the diagonal line, moves to bluer colors and enters the PSB region from the top right, and then pivots to continue moving to redder colors.
They find that the timing ($z_{\rm quench} \sim 2$ vs.~$z_{\rm quench} \sim 1$) and the speed of quenching both affect this model track in subtle ways. 

However, these models are built on simplified assumptions for both the galaxy star formation history and the dust attenuation curve.
We therefore employ our cosmological simulations in order to assess the role that quenching timescale may have on the UVJ color evolution of redshift $z=1-2$ galaxies.
To do this, we first compute quenching times following the definition of \citet{rodriguezmontero_mergers_2019}: we compute the quenching timescale $\tau_q$, in Gyr, as the time it takes the galaxy to go from the ``star-forming threshold'' $\mathrm{sSFR} > t_H^{-1}$ to the ``quenched threshold'' $\mathrm{sSFR} < 0.2~t_H^{-1}$, where $t_H$ is the age of the universe at that epoch. 
We refer to the times at which a galaxy crosses the star-forming and quenched thresholds as $t_{\rm start}$ and $t_q$, respectively.
For a galaxy to be quenched, we additionally impose the requirement that it remain below the star-forming threshold for an additional $0.2~t_q$ after quenching. 
By this definition, $\tau_q$ represents the time it takes a galaxy to cross the green valley, comparable to the distinction between fast and slow provided by \citet{carnall_vandels_2019}. 
\citet{rodriguezmontero_mergers_2019} found that \simba\ galaxies under this definition naturally divide into ``fast'' ($\tau_q \sim 0.01 t_q$) and ``slow'' ($\tau_q \sim 0.1 t_q$) quenching modes.

Figure~\ref{fig:time_ev_tau_q} shows how galaxy evolution in UVJ space depends on $\tau_q$. 
We explore this question in two ways: first, by computing the median track in UVJ space for fast vs.~slow quenching galaxies (middle panel), and second, by computing the UVJ colors of each galaxy at the time of quenching $t_q$ (right panel).
The left panel of Figure~\ref{fig:time_ev_tau_q} shows SFHs, scaled by $t_H$ to fit the quenching thresholds and with the $x$-axis centered on $t_q$. 

\begin{figure*}
	\centering
	\includegraphics[width=\linewidth]{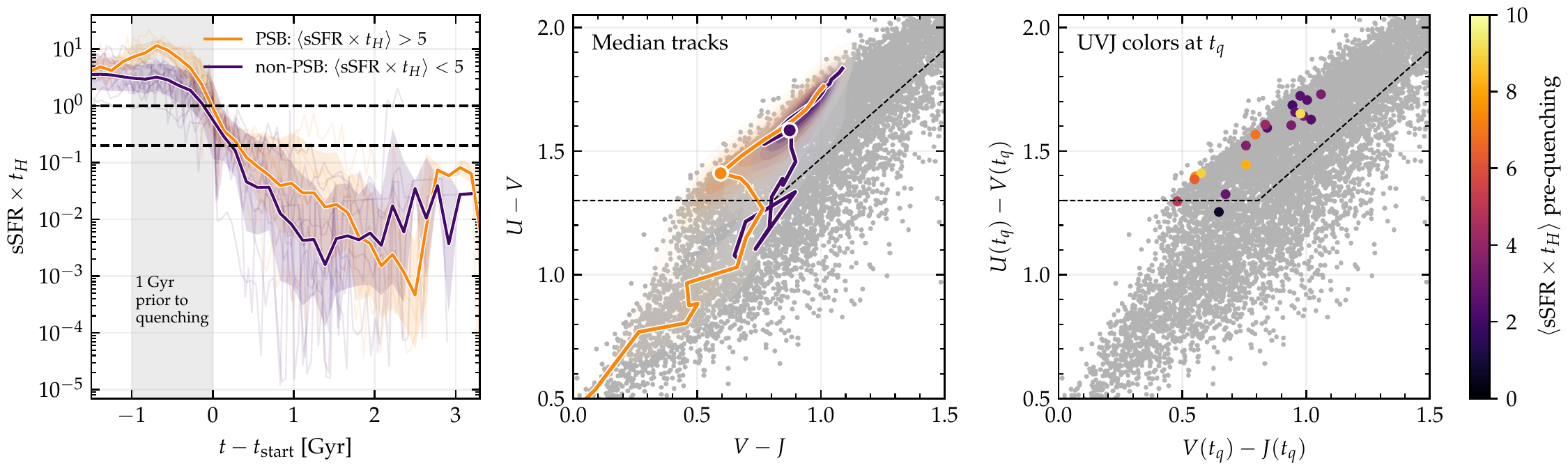}
	\caption{Evolutionary tracks in UVJ space as a function of the mean sSFR over the 1 Gyr prior to the onset of quenching, $\langle {\rm sSFR} \times t_H \rangle$. Note that in the left panel, the $x$-axis is now centered on $t_{\rm start}$, the time at which the galaxy first leaves the star-forming threshold, and the gray shaded region indicates the 1 Gyr timespan on which we average the sSFR. We see a distinct difference in the median UVJ evolutionary track for galaxies that quench immediately post-starburst.}	
	\label{fig:time_ev_sSFR}
\end{figure*}

We compute median tracks by taking the median $U-V$ and $V-J$ colors for our sample in bins of $t-t_q$. 
We split our sample into fast- and slow-quenching at $\log_{10}(\tau_q/t_H) = -1.5$ or roughly $\tau_q \approx 0.03 t_H$.
We additionally plot kernel density estimate (KDE) contours showing the distribution of fast- and slow-quenching UVJ colors at all timesteps.
These contours capture the dispersion of UVJ trajectories underlying the median track and are intended to highlight the diversity.
It is clear from Figure~\ref{fig:time_ev_tau_q} that there is not a distinct difference between the UVJ evolution of fast vs.~slow quenching galaxies. 
In fact, if anything, slow-quenching galaxies preferentially enter the quiescent region from the bottom left, in direct conflict with what is inferred from observations.
Nevertheless, despite substantial differences in $\tau_q$, the median tracks and the UVJ colors at $t_q$ do not indicate a clear preference for fast- vs.~slow-quenching galaxies to enter the quenched region from different locations.

The definition of $\tau_q$ as the time it takes a galaxy to cross the green valley is not the only way to assess fast vs.~slow quenching. 
In order to provide a more direct comparison to the results of \citet{belli_mosfire_2019}, we also fit a simple exponentially-declining model to the SFHs of each simulated galaxy.
We consider only a limited portion of each SFH, starting at the time of peak SFR prior to quenching and ending 1 Gyr after quenching. 
We fit a decaying exponential model using 
\begin{equation}
  {\rm SFR}~\propto~e^{-({t-t_0})/{\tau_{\rm exp}}}
\end{equation}
where $t_0$ is the time of peak SFR and $\tau_{\rm exp}$ is the decay time, the parameter to be fit. 

Figure~\ref{fig:time_ev_tau_exp} shows how galaxy evolution in UVJ space depends on $\tau_{\rm exp}$. 
Here, we split the sample into fast- and slow-quenching at $\tau_{\rm exp} = 400$ Myr in order to correspond roughly to the $100$ Myr and $1$ Gyr model tracks presented by \citet{belli_mosfire_2019}.
 As with $\tau_q$, we do not see clear evidence for distinct UVJ evolutionary tracks based on the quenching timescale: the median tracks are nearly identical and galaxies with different $\tau_{\rm exp}$ seem to enter the quenched region from similar locations. 
This is in distinct contrast to the results of \citet{belli_mosfire_2019} and \citet{carnall_vandels_2019}, and implies that the primary factors driving the evolution of UVJ colors in \simba\ are not strongly correlated with the quenching timescale.

\subsection{Post-starburst galaxies}\label{sec:psbs}

Despite the lack of distinct evolutionary tracks for fast vs.~slow quenching, the observational evidence for the clustering of post-starburst galaxies in the lower-left of the quenched region in UVJ space implies that these galaxies must follow a unique evolutionary track. 
We have shown that this is indeed the region where our youngest quenched galaxies tend to lie (see Figure~\ref{fig:uvj_age}). 

While PSB galaxies are typically associated with fast quenching timescales, this correlation may not be universal. 
In the dual-origin model for PSBs presented by \citet{wild_evolution_2016}, at low-redshift ($z\lesssim 1$), PSBs are formed by the {\it rapid} quenching of normal star-forming galaxies, while at high-redshift ($z \gtrsim 2$), PSBs are formed by a period of intense starburst and subsequent quenching. 
That is, at $2.6 \lesssim z < 1$, we may not expect PSBs to necessarily show universally rapid quenching timescales but rather be characterized by intense starburst prior to quenching.
Therefore, we classify galaxies by computing their mean sSFR (scaled by $t_H$) in the 1 Gyr prior to the onset of quenching. 
We write this quantity as $\langle {\rm sSFR}\times t_H\rangle_{\rm pre-quenching}$ or simply $\langle {\rm sSFR}\times t_H\rangle$. 
Since the sSFR can be interpreted as the inverse of the stellar mass doubling time, a value of $\langle {\rm sSFR}\times t_H\rangle > 5$ would indicate that the stellar mass could double in less than one-fifth of a Hubble time, or $\sim 1$ Gyr at $z\sim 1$.

Figure~\ref{fig:time_ev_sSFR} shows the evolution of UVJ colors as a function of $\langle {\rm sSFR}\times t_H\rangle$. 
In the left panel, we again plot the SFHs, but this time with the $x$-axis centered on $t_{\rm start}$, time at which quenching began, in order to highlight the 1 Gyr timespan prior to quenching on which we average the sSFR. 
We split the sample into two groups, ``PSBs'' with $\langle {\rm sSFR}\times t_H\rangle > 5$ and ``non-PSBs'' that don't satisfy this criteria.
This definition yields 7 PSBs and 13 non-PSBs.
The median SFHs for these two groups are distinctly different, with the PSB group showing an extreme peak in the SFH prior to quenching but taking longer to reach lower sSFRs after quenching. 
We also see a distinct difference in the median UVJ diagram tracks. 
PSB galaxies, which quench following a starburst, typically veer towards the quenched region early and enter from the bottom left. 
In contrast, non-PSB galaxies move to redder colors in the star-forming region before entering the quenched region along the diagonal boundary.
While there is significant diversity in the UVJ evolutionary pathways underlying the median track, the PSB population shows a particularly high density just outside the quenched region on the bottom left.
We can additionally see this effect on the right panel: most of the galaxies entering the quenched region from the bottom left are PSBs, while those entering from the right are not.

\begin{figure}
    \centering
    \includegraphics[width=\linewidth]{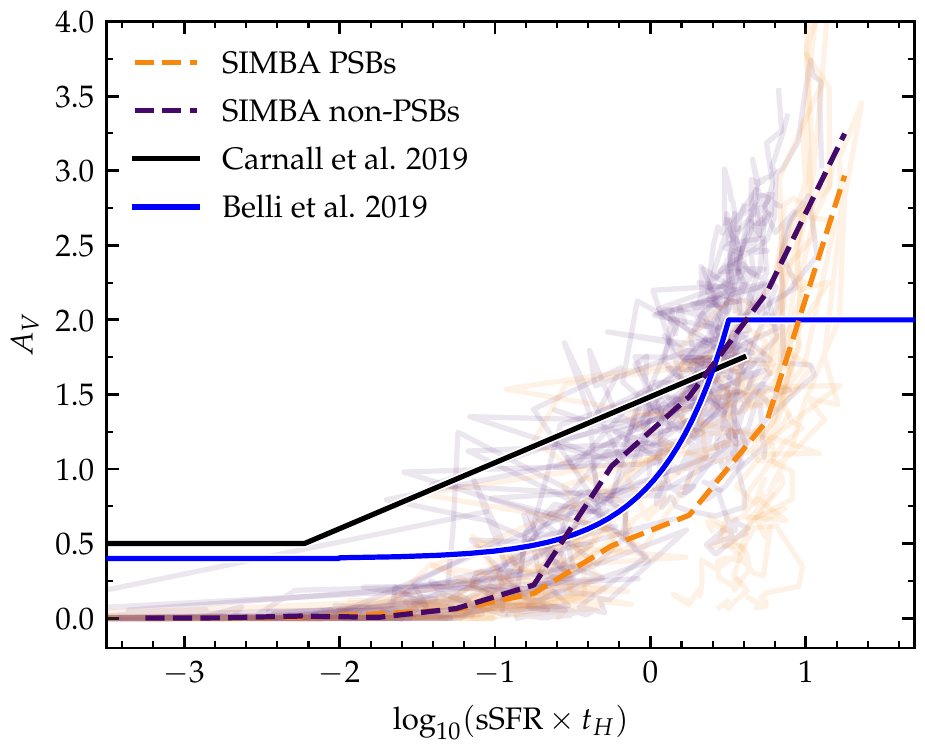}
    \caption{Dust attenuation $A_V$ vs. $\log_{10}({\rm sSFR}\times t_H)$ for SIMBA PSB galaxies (orange) and non-PSBs (purple). For both populations, we show both the median $A_V$ in bins of $\log_{10}({\rm sSFR}\times t_H)$ as well as tracks for individual galaxies. The solid black line shows the ansatz used by \citet{carnall_vandels_2019} to infer model tracks on the UVJ diagram. The solid blue line shows the ansatz used by \citet{belli_mosfire_2019}. Though not perfect, our results in general more closely match the \citeauthor{belli_mosfire_2019} ansatz. Furthermore, our PSB galaxies drop in $A_V$ more quickly, or at higher sSFR, than the non-PSB population.}
    \label{fig:carnall_belli}
\end{figure}

We see that PSBs follow a unique evolutionary pathway in UVJ space, but what physical mechanism causes this divergence from the non-PSB galaxies? 
As it turns out, PSBs are driven into this region by a rapid loss of dust as they quench. 
In the \simba\ model, dust can be destroyed in star-formation, primarily due to astration in the hot environments near stars and, to a lesser degree, thermal sputtering in the ISM \citep{li_dusttogas_2019}\footnote{Note: these simulations do not include shattering as a dust destruction process \citep[e.g.][]{li_origin_2021}.}. 
PSBs in \simba\ thus lose most of their dust during starburst, and converge upon a dust-free model track in UVJ space earlier than non-PSB galaxies (i.e., before fully quenching). 
For example, we see in Figure~\ref{fig:time_ev_examples} that galaxy 36 experiences a rapid drop in dust attenuation during starburst, such that it has $A_V \sim 0.5$ by the time it begins quenching.
By contrast, galaxy 6, which does not experience a starburst, has $A_V \sim 1$ at the start of quenching. 

Accurately assessing the rate at which dust is lost during quenching is critical in deriving UVJ evolutionary tracks from observations. 
This is typically done by assuming a relationship between the SFR and the dust attenuation $A_V$ to model the change in colors with decreasing SFR.
This is the method employed by both \citet{belli_mosfire_2019} and \citet{carnall_vandels_2019}, though the model tracks produced by the two authors differ significantly. 
This is largely because \citet{belli_mosfire_2019} have no data to constrain this relationship and simply assume that $A_V$ falls linearly with SFR as galaxies quench. 
By contrast, \citet{carnall_vandels_2019} fit a linear relationship between $A_V$ and $\log({\rm nSFR})$ for a sample of green valley galaxies.

Figure~\ref{fig:carnall_belli} shows how these two models compare to our results for the relationship between $A_V$ and sSFR.\footnote{To plot the ansatz of \citet{carnall_vandels_2019}, we convert the equation provided from $\log{({\rm nSFR})}$ to $\log{({\rm sSFR}\times t_H)}$ using a linear fit derived from \simba\ data. To plot the ansatz of \citet{belli_mosfire_2019}, we assume that galaxies start with $\log({\rm sSFR}\times t_H) = 0.5$ and $A_V=2$, and that $A_V$ scales with ${\rm sSFR}\times t_H$ until settling at a constant value of $A_V = 0.4$ at $\log({\rm sSFR}\times t_H) = -2$. While differing definitions prohibit perfect comparison between these assumptions and our results, they remain useful for qualitative comparison.}
We show the relationship between $A_V$ and $\log({\rm sSFR}\times t_H)$ for both our PSB galaxies (orange) and non-PSBs (purple). 
First, we see that our results more closely match the \citeauthor{belli_mosfire_2019} ansatz, where $A_V$ drops following the SFR. 
Furthermore, we note that the PSB galaxies drop to lower $A_V$ at the same sSFR than the non-PSB population. 
This reflects the rapid destruction of dust during starburst, and is a major reason we see PSBs follow the UVJ pathway they do, despite not necessarily quenching rapidly.

\subsection{Time Evolution Summary}

These results, in general, support the interpretations of recent observations \citep[e.g.][]{barro_candels_2014, belli_mosfire_2019, carnall_vandels_2019, suess_dissecting_2021} that there are, to first order, two quenching pathways in UVJ space: one that enters the quenched region at some point along the diagonal boundary and another that enters from the bottom left. 
While we don't observe a difference in UVJ evolution based on the quenching timescale, we do see that the latter pathway is primarily associated with PSBs.\footnote{This result---that PSBs in \simba\ do not necessarily quench rapidly, but do follow the PSB evolutionary track inferred from observations---may explain why we observe an overdensity of transition galaxies just outside the quenched region on the bottom left: PSBs in \simba\ quench slowly relative to observations and spend more time in this transition region.}
However, we caution that these median tracks are indeed the median of a diverse array of UVJ evolutionary paths. 
As we show in Figure~\ref{fig:time_ev_examples}, two galaxies that follow similar evolution in general, or enter the quenched region in similar ways, may trace significantly different paths as they move somewhat chaotically through the blue cloud. 
While this may be due simply to the discreteness of our simulation snapshots and the dependence of our radiative transfer on the evolving geometry of the galaxy, we nevertheless caution that the evolution of galaxies in color-color space is sensitive to more than just SFR and $A_V$.

The lack of a clear relationship between the quenching timescale and the formation history of our model galaxies holds implications for observations. 
In fact, recent resolved HST observations of galaxies at $z\gtrsim 1$ find that galaxies that form late experience fast quenching in their centers whereas early formation correlates with slow/uniform quenching \citep{Akhshik_2022}. Thus, it may be critical to explicitly consider the quenching within the central $1$ kpc when connecting  quenching timescales to the global formation histories in future theoretical work.

\section{Discussion: Comparison to Other Models}\label{sec:discussion}

In this section, we compare the results of our simulations to theoretical models in the literature that have attempted to understand the observed UVJ diagram. 

\subsection{Overview}

We compare to three different simulation campaigns that have studied the UVJ diagram in the content of galaxy evolution: \citet{dave_mufasa_2017}, \citet{donnari_star_2019}, and \citet{roebuck_simulations_2019}.
\citet{dave_mufasa_2017} and \citet{donnari_star_2019} employed cosmological galaxy evolution simulations (similar to those studied here), while \citet{roebuck_simulations_2019} focused on idealized galaxy models.
The former papers simulated synthetic colors via LOS ray-tracing models, while \citet{roebuck_simulations_2019} used bona fide radiative transfer calculations as in our work.
In what follows, we compare both the quenching models (and impact on the separation between star-forming and quenched galaxies) and the impact of assumed/modeled dust attenuation laws in these works on the modeled colors.

\subsection{Galaxy quenching models}

Critical to computational studies of galaxy quenching is the underlying physical model responsible for quenching massive galaxies. 
In recent years, an AGN-driven scenario for quenching in massive galaxies has gained traction \citep[see e.g.][]{dubois_agndriven_2013}.
Indeed, the inclusion of AGN feedback in \simba\ is one of the major differences between it and its predecessor simulation, \textsc{mufasa} \citep{dave_mufasa_2016}.
While \textsc{mufasa} used a phenomenological model for quenching, which prevents gas from cooling onto galaxies in halos above a certain redshift-evolving halo mass threshold, \simba\ allows galaxies to quench naturally based on subgrid models for black hole feedback. 
\citet{dave_simba_2019} show that these models for AGN feedback are primarily responsible for quenching massive galaxies, with jet-mode feedback dominating but X-ray feedback playing a subtle but important role.

The quenching model used in IllustrisTNG is broadly similar to \simba\, in which quenching is driven primarily by kinetic AGN feedback.
That said, there are some key differences in the two models. 
In particular, \simba\ does not vary the direction of AGN jets as IllustrisTNG does. 
Additionally, \simba\ employs bipolar kinetic AGN feedback at all Eddington ratios, while IllustrisTNG employs spherical thermal feedback at high ratios and kinetic feedback at low ratios. 
The differences between the \simba\ an IllustrisTNG AGN feedback implementation, while subtle, have been shown to play a role in determining the cold gas content of SFGs, particularly at high redshift \citep{dave_galaxy_2020}.

The slight overpopulation of the green valley in \simba\ (which contributes to the lack of a clear UVJ bimodality) is dominated by galaxies with $10 < \log M_*/{\rm M}_\odot < 11$ \citep[][their Figure 6]{dave_simba_2019}. 
In contrast, the distribution of galaxy SFRs in TNG does not show overpopulation of the green valley in this same mass range \citep[][their Figure 8]{donnari_star_2019}. 
These differences in the distribution of SFRs between the two simulations are likely driven by differences in the feedback implementations: TNG feedback randomizes the jet direction and is able to expel the ISM in the low-$f_{\rm edd}$ mode, while \simba\ assumes bipolar jets that are decoupled until beyond the ISM.
This is likely an important reason why IllustrisTNG produces a clear color bimodality on the UVJ diagram whereas \simba\ does not. 

\newpage
\subsection{Dust attenuation models}

The treatment of dust in the models of \citet{dave_mufasa_2017}, \citet{donnari_star_2019}, and \citet{roebuck_simulations_2019} all differ from each other and from this work in key ways. 
\citet{dave_mufasa_2017} derive a dust attenuation curve for each galaxy using the ray-tracing package \textsc{loser}\footnote{\url{https://pyloser.readthedocs.io/en/latest/}}, which acts as a computationally inexpensive alternative to dust radiative transfer. 
They use a redshift-dependent dust-to-metals ratio and a \citet{cardelli_relationship_1989} Milky Way extinction law\footnote{\textsc{pyloser} is now included by default in \textsc{caesar}.}. 
Meanwhile, the dust model used by \citet{donnari_star_2019} \citep[described in detail in][]{nelson_first_2018}, includes the empirical model of \citet{charlot_simple_2000} and additionally models dust scattering analytically following \citet{calzetti_dust_1994} and dust absorption following \citet{cardelli_relationship_1989} with a redshift and metallicity-dependent dust-to-gas ratio. 
Finally, \citet{roebuck_simulations_2019} use idealized simulations and radiative transfer to determine galaxy colors, though they assume a constant dust-to-metals ratio of $0.4$ \citep[as opposed to our on-the-fly model for dust evolution from][]{li_dusttogas_2019}. 

A major difference between the models used in this work and in those used by \citet{dave_mufasa_2017} and \citet{donnari_star_2019} is the use of dust radiative transfer vs.~LOS extinction. 
The inclusion of full 3D dust radiative transfer is important in capturing the variation of the dust attenuation curve, and indeed, \citet{roebuck_simulations_2019} also find significant variation in the attenuation curve which drives galaxy locations in UVJ space. 
This is likely a significant reason that our models produce a larger spread of UVJ colors for SFGs---including populating the dusty star-forming region---than those of \citet{dave_mufasa_2017} or \citet{donnari_star_2019}, which generally employ attenuation curves with fixed shapes.

Notably, while our model succeeds in reproducing the observed population of highly dust-reddened galaxies, there remain tensions with observations. 
In particular, as discussed in Section~\ref{sec:uvj_av}, the attenuation curve shapes of our model galaxies differ from the typically assumed \citet{calzetti_dust_2000} curve, and this difference is more significant at higher masses. 
We attribute this to the increasingly complex star-dust geometry in higher-mass galaxies, which may be related to the quenching model.
As dust in \simba\ is advected passively with gas elements, the star-dust geometry would be impacted by any process that alters the spatial distribution of the gas. 
The AGN feedback model in \simba\ does just that: \citet{borrow_cosmological_2020} show that jet-mode AGN feedback is capable of transferring galaxy baryons great distances, in some cases several Mpc.
Therefore, while the radiative transfer models employed in this work are certainly more robust than LOS extinction models, they are dependent on the somewhat unconstrained evolution of the 3D distribution of gas at the epoch of quenching.
Recent simulations such as those performed by \citet{li_origin_2021}, which decouple gas and dust, may help to address this issue.

\section{Conclusions}\label{sec:conclusions}

In this work, we have studied the evolution of galaxies on the UVJ diagram using the \simba\ simulations and \pd\ 3D dust radiative transfer. 
Our main conclusions are summarized as follows: 
\begin{enumerate}
	\item The \simba\ dust model in combination with \pd\ dust radiative transfer broadly reproduces the observed distribution of galaxies on the UVJ diagram at $z=2$ and $z=1$. In particular, we reproduce:
	\begin{itemize}
		\item the clustering of galaxies into star-forming and quiescent regions;
		\item the relationship between $V-J$ color and $A_V$ for SFGs, including the population of extremely dust-reddened galaxies at the top right of the diagram;
		\item the diagonal gradient in stellar age for quiescent galaxies. 
	\end{itemize} 
	\item However, we fail to reproduce observations in several key ways: 
	\begin{itemize}
		\item We do not reproduce a clear bimodality in the number density of galaxies in UVJ space, likely due to the overpopulation of the green valley in \simba\ and the dust attenuation curve shapes.  
		\item We populate the dusty star-forming region primarily with low-mass ($\log M_*/\Msun < 10.5$), rather than higher-mass galaxies. These low-mass, high $A_V$ galaxies are not typically found in observational surveys and may be due to issues with the \simba\ dust evolution model.
	\end{itemize}
	\item We find that the assumption of a universal, \citet{calzetti_dust_2000} dust attenuation law can lead to bias in the inferred SFRs in the star-forming region of UVJ space, with SFGs near the quiescent region having their SFRs underestimated by as much as $0.5$ dex. We caution that trends in UVJ space may be exaggerated by the assumption of a universal attenuation law.
	\item In contrast to what is typically inferred from observations, we find little correlation between the quenching timescale and the pathway a galaxy follows in UVJ space as it quenches. 
	Instead, we show that the evolution of galaxies in UVJ space is driven primarily by the intensity of its star-formation in the 1 Gyr prior to the onset of quenching.
	Galaxies that experience a burst of star-formation prior to quenching veer to the left edge of the blue cloud and enter the quenched region from the bottom left. 
	Galaxies that do not experience such a burst in star-formation enter the quenched region along the diagonal boundary. 
\end{enumerate}
Interpretation of our results is limited by the extent to which we fail to reproduce observed distribution of galaxies in UVJ space. 
The fact that simulations---even those employing an explicit dust model and 3D radiative transfer---still cannot perfectly reproduce observations of color-color diagrams at high redshift highlights the need for further work modeling the relationship between dust attenuation, star-formation, and morphological transition for galaxies in the process of quenching. 
Central to these questions is the relationship between dust geometry and the dust attenuation law \citep[see][]{narayanan_theory_2018}, the evolution of galactic dust during quenching \citep[see][]{whitaker_2021}, and the morphological evolution of galaxies during quenching. 
Future advancements in galaxy dust modeling \citep[e.g.][]{li_origin_2021} will allow us to more rigorously explore the evolution of dust properties, and such analysis will be key for improving our interpretation of observable properties of high-redshift galaxies. 

\acknowledgements

This work was supported by NSF under grant AST-1908137 and REU-1851954. H.B.A. acknowledges the feedback and support provided by the faculty and students involved in the 2020 UF REU program. 
K.E.W. wishes to acknowledge funding from the Alfred P. Sloan Foundation.
R.F. acknowledges financial support from the Swiss National Science Foundation (grant no 194814).
This work was initiated at the Aspen Center for Physics, which is supported by NSF grant PHY-1607611.
Much of this work was conducted on the ancestral land of the Meskwaki, Sauk, and Ioway Peoples.

\software{\textsc{Python}, \textsc{numpy} \citep{walt_numpy_2011}, \textsc{matplotlib} \citep{hunter_matplotlib_2007},  \textsc{powerderday} \citep{narayanan_powderday_2021}, \textsc{yt} \citep{turk_yt_2011}, \textsc{hyperion} \citep{robitaille_hyperion_2011}, \textsc{fsps} \citep{conroy_propagation_2010,conroy_propagation_2010a}, \textsc{gizmo} \citep{hopkins_new_2015}.}

\newpage
\appendix
\restartappendixnumbering
\section{UVJ colors over time for the full sample}\label{appendix:timeev}

\begin{figure*}[h!]
    \centering
    \includegraphics[width=\linewidth]{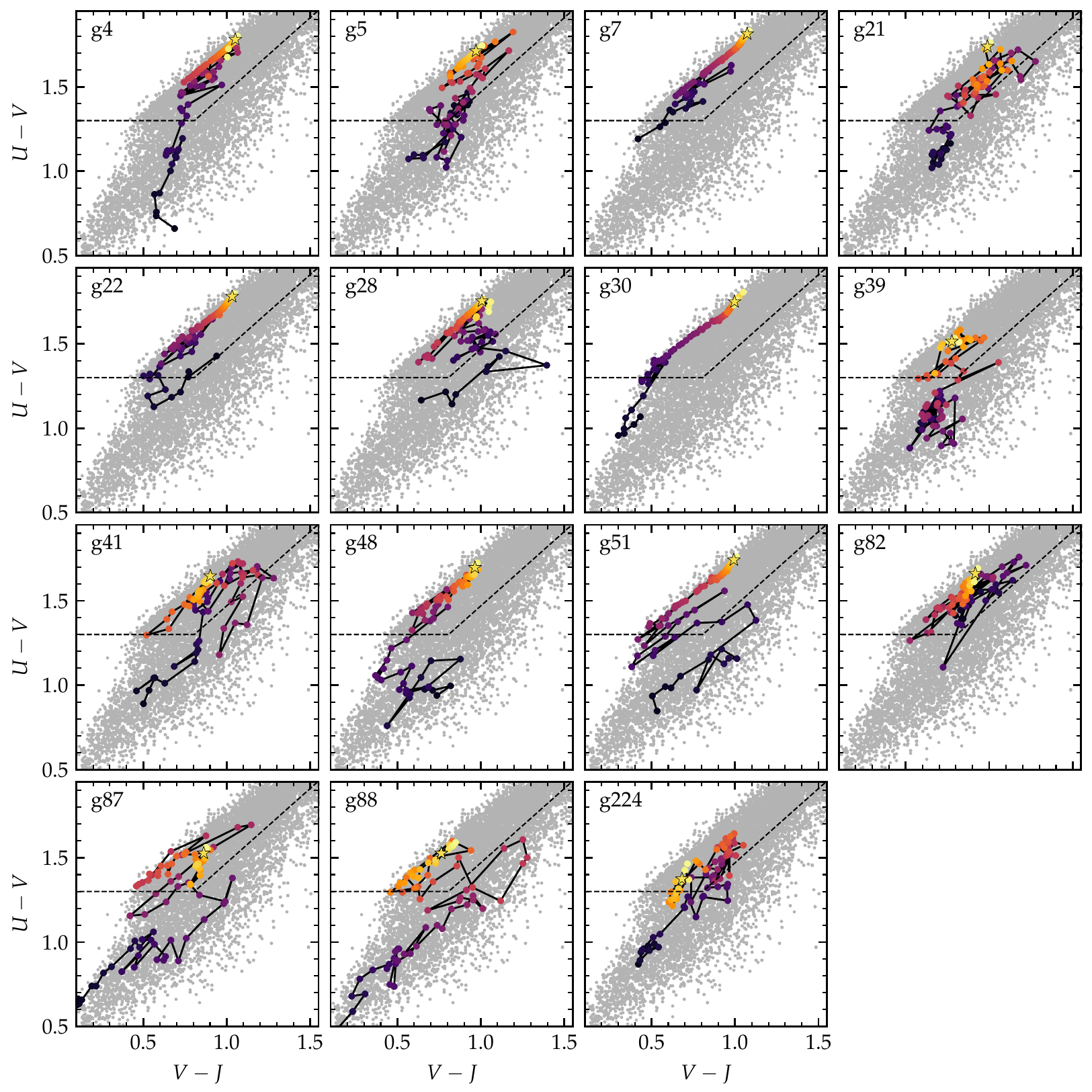}
    \caption{Evolution of UVJ colors from $z\sim 2.6$--$1$ for the rest of the galaxies in our sample, not pictured in Figure~\ref{fig:time_ev_examples}. Points are colored according to the age of the universe at that redshift, from $2.4$ to $6.3$ Gyr. The dashed line shows the \citet{williams_detection_2009} UVJ selection criteria.}
    \label{fig:time_ev_examples_all}
\end{figure*}

Figure~\ref{fig:time_ev_examples_all} shows the evolution of UVJ colors from $z\sim 2.6$ to $1$ for the remaining galaxies in our sample, not pictured in Figure~\ref{fig:time_ev_examples}. 

\restartappendixnumbering
\section{Mock observational noise}\label{appendix:noise}

\begin{figure*}[h!]
    \centering
    \includegraphics[width=0.8\linewidth]{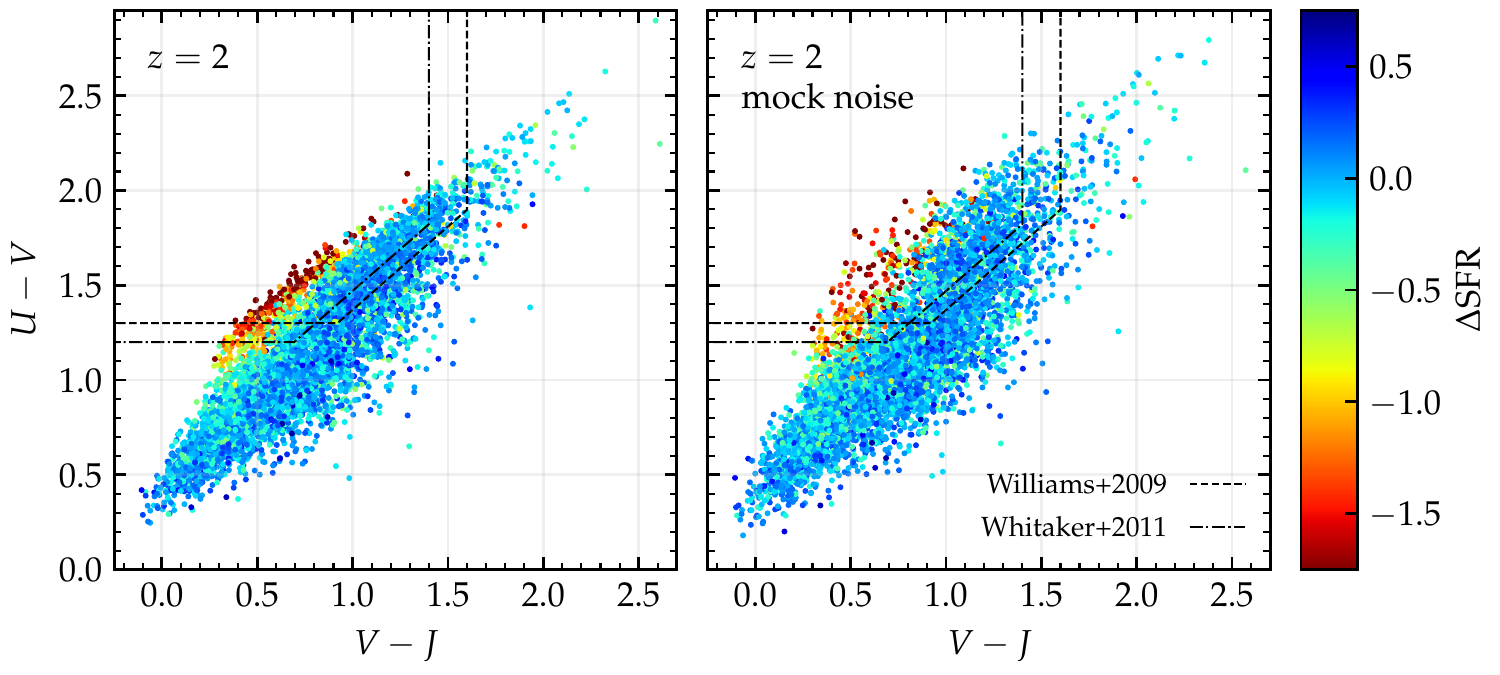}
    \caption{The UVJ diagram at $z=2$, both with (right) and without (left) the addition of mock observational noise.
    Points are colored by $\Delta$SFR.}
    \label{fig:noise}
\end{figure*}

Figure~\ref{fig:noise} shows the \simba+\pd\ model UVJ diagram both with and without the inclusion of mock observational noise.
We simulate the observational noise in the form of uncertainty in the redshift, e.g. from photometric redshift measurements in large surveys. 
Specifically, we apply an uncertainty of $\sigma_z/(1+z) = 0.05$ for ``red'' galaxies (with $U-V > 1.3$) and $\sigma_z(1+z) = 0.025$ for ``blue'' galaxies (with $U-V < 1.3$). 
We randomly select $z_{\rm phot}$ from a normal distribution ${\rm Norm}(2, \sigma_z)$ and calculate the rest-frame magnitudes from the shifted SED, assuming $z_{\rm phot}$. 
We apply an increased uncertainty for red vs.~blue galaxies following \citet[Figure 21]{whitaker_newfirm_2011}. 

\bibliographystyle{aasjournal} 
\bibliography{UVJ_paper.bib}

\end{document}